\documentclass[10pt]{article}

\usepackage[margin=1in]{geometry}
\usepackage{mathpazo}
\usepackage[backend=biber,natbib=true,style=apa]{biblatex}
\usepackage{xltabular}
\usepackage{authblk}

\usepackage{microtype}
\usepackage{graphicx}
\usepackage{subcaption}
\usepackage{caption}
\usepackage{todonotes}

\usepackage{booktabs} 

\usepackage{bbm}
\usepackage[labelfont=bf]{caption}

\usepackage{titlesec}

\usepackage{psfrag}
\usepackage{amsmath}
\usepackage{amsfonts}
\usepackage{verbatim}
\usepackage{mathrsfs}
\usepackage{amssymb}
\usepackage{pifont}

\usepackage{multirow}
\usepackage{longtable}
\usepackage{listings}
\usepackage{tabularx}
\usepackage{subcaption}

\usepackage{color, colortbl}
\usepackage{xspace}
\usepackage{xr}
\usepackage{lineno}
\usepackage{fancyhdr}
\usepackage[english]{babel} 
\usepackage{threeparttable}
\usepackage{lipsum} 
\usepackage{float}
\usepackage{minted}
\usepackage{array} 
\usepackage{xcolor}  
\usepackage{soul}    
\sethlcolor{yellow}
\usepackage{listings}
\usepackage{subcaption}
\usepackage{tablefootnote}
\usepackage{array}
\usepackage{xcolor}
\usepackage{colortbl}
\usepackage{booktabs}
\usepackage{multirow} 
\usepackage{tabularx, makecell, booktabs}
\usepackage[none]{hyphenat}
\titleformat{\section}[block]{\normalfont\Large\bfseries}{}{0pt}{}
\titleformat{\subsection}[block]{\normalfont\large\bfseries}{}{0pt}{}
\titleformat{\subsubsection}[block]{\normalfont\normalsize\bfseries}{}{0pt}{}

\title{Evaluating 21st-Century Competencies in Postsecondary Curricula with Large Language Models: Performance Benchmarking and Reasoning-Based Prompting Strategies}

\author[1]{Zhen Xu}
\author[1]{Xin Guan}
\author[1]{Chenxi Shi}
\author[2]{Qinhao Chen}
\author[1,3*]{Renzhe Yu}
\affil[1]{Teachers College, Columbia University}
\affil[2]{Graduate School of Arts and Sciences, Columbia University}
\affil[3]{Data Science Institute, Columbia University}
\affil[*]{Correspondence: \texttt{renzheyu@tc.columbia.edu}}

\date{}
\makeatletter
\renewcommand{\@maketitle}{%
  \newpage
  \null
  \vskip -2.0em%
  \begin{center}%
    {\small\textcolor{red}{Due to appear in Journal of Learning Analytics Special Section on Advancing 21st-century Professional Competencies with Learning Analytics in the Age of Generative AI}\par}%
    \vskip 0.8em%
    {\Large \bfseries \@title \par}%
    \vskip 1.0em%
    {\large
      \lineskip .5em%
      \begin{tabular}[t]{c}%
        \@author
      \end{tabular}\par}%
  \end{center}%
  \par
  \vskip 1.2em%
}
\makeatother

\begin{document}
\maketitle
\begin{abstract}
The growing emphasis on 21st-century competencies in postsecondary education, intensified by the transformative impact of generative AI on the economy and society, underscores the urgent need to evaluate how they are embedded in curricula and how effectively academic programs align with evolving workforce and societal demands. Curricular Analytics (CA), particularly recent advancements powered by generative AI, offer a promising data-driven approach to this challenge. However, the analysis of 21st-century competencies requires pedagogical reasoning beyond surface-level information retrieval, and the capabilities of large language models (LLMs) in this context remain underexplored. In this study, we extend prior research on curricular analytics of 21st-century competencies across a broader range of curriculum documents, competency frameworks, and models. Using 7,600 manually annotated curriculum-competency alignment scores (38 competencies and 200 courses across five curriculum document types), we evaluate the informativeness of different curriculum document sources, benchmark the performance of general-purpose LLMs on mapping curricula to competencies, and analyze error patterns. We further introduce a reasoning-based prompting strategy, Curricular CoT, to strengthen LLMs’ pedagogical reasoning. Our results show that detailed instructional activity descriptions are the most informative type of curriculum document for competency analytics. Open-weight LLMs achieve accuracy comparable to proprietary models on coarse-grained tasks, demonstrating their scalability and cost-effectiveness for institutional use. However, no model reaches human-level precision in fine-grained pedagogical reasoning. Our proposed Curricular CoT yields modest improvements by reducing bias in instructional keyword inference and improving the detection of nuanced pedagogical evidence in long text. Together, these findings highlight the untapped potential of institutional curriculum documents and provide an empirical foundation for advancing AI-driven curricular analytics.
\end{abstract}

\section{Introduction}
Globalization and rapid technological advancements, particularly the recent emergence of generative artificial intelligence (GenAI), are profoundly transforming the labor market and the broader economic landscape. As AI automates routine and repetitive tasks, human labor is expected to shift toward roles that require higher-order cognitive skills, creativity, and social-emotional intelligence ~\citep{oecd2023employment}. In this context, foundational knowledge and technical skills are no longer sufficient for future workforce readiness. Employers now place growing importance on a broad set of transferable, domain-agnostic competencies, commonly referred to as 21st-century competencies, such as critical thinking, complex problem-solving, collaboration, communication, adaptability, and digital literacy ~\citep{oecd2023}. Recognizing these global shifts, policymakers and organizations worldwide are underscoring the urgency of equipping future generations with 21st-century competencies to meet the demands of an increasingly digital and AI-driven economy ~\citep{oecd2023employment,ellingrud2023generative,wef2025}.

Correspondingly, these shifts carry profound implications for teaching and learning, especially in postsecondary education, where academic preparation directly connects to workforce readiness. Postsecondary education institutions are increasingly expected to integrate and foster higher-order, interdisciplinary 21st-century competencies within their traditional subject-based curricula ~\citep{unesco2023genai,oecd2023digital, nea2024ai}. In response, many countries have introduced new teaching and learning frameworks to guide systematic reforms and embed these competencies into national learning outcomes, encouraging institutions to align with evolving societal and workforce needs ~\citep{used2023ai,ukgenerative}. However, institutions still face challenges in translating these policy expectations into practice ~\citep{oecd2023, buckingham2016learning}, particularly in redesigning curricula and instruction. Effective reform requires faculty training and pedagogical support, digital and technical infrastructure, and flexible institutional policies—resources that many institutions lack or possess unevenly. Therefore, examining curriculum design and implementation at scale is critical to understanding how institutions are responding to the evolving demands of 21st-century education and adapting to future workforce preparation.

Traditional curricular analytics and evaluation have long relied on experts' manual analysis, which is labor-intensive and difficult to scale. With ongoing advancements in postsecondary education data infrastructure and analytical techniques, more institutions and stakeholders are seeking automated, data-driven approaches. In this context, curricular analytics is a promising approach that leverages computational methods to support data-driven curriculum decision-making, program improvement, and institutional evaluation ~\citep{hilliger2020design}. By leveraging digital records, curricular analytics enables scalable and systematic analysis of various dimensions of curriculum and instruction, such as learning objectives, pedagogical design and practices, learning outcomes, and their alignment with workforce competencies ~\citep{pistilli2017guiding,chou2015open,hilliger2024curriculum}.  

However, the analysis of 21st-century competencies in postsecondary curricula remains at an early stage of exploration, primarily due to technical and data limitations. Technically, it requires pedagogically informed reasoning to interpret learning activities and instructional design, shifting the task beyond traditional NLP approaches (e.g., rule-based matching or embedding-based alignment) toward pedagogical reasoning. Although supervised machine learning offers partial solutions, it depends on large, manually annotated datasets that are scarce in postsecondary education, limiting scalability. Moreover, from a data quality perspective, such analysis also requires detailed, high-quality curriculum documents, as overly general materials often lack sufficient information. As a result, despite institutions maintaining extensive collections of curriculum documents, their potential to generate actionable insights into competency development remains largely untapped.

Recent advancements in generative AI, particularly large language models (LLMs), show promise for addressing these challenges. Some recent studies evaluate LLMs’ ability to analyze higher-order interdisciplinary competencies in domains such as business and information science ~\citep{jayalath2025scaling,jovanovic2025curriculum} and suggest that LLMs show initial promise. However, other work highlights persistent limitations, especially for tasks requiring pedagogically grounded reasoning and interpretation of instructional design ~\citep{zamecnik2024mapping,xu2025course}. Without careful prompt design, fine-tuning, or contextual adaptation, LLMs struggle to reliably perform complex pedagogical reasoning tasks. Therefore, fully realizing the potential of LLMs for assessing 21st-century competencies in postsecondary curricula requires systematic cross-contextual evaluation to identify when LLMs are effective and how their pedagogical reasoning and contextual inference can be improved.

This study aims to advance AI-assisted curricular analytics by systematically examining the capabilities of LLMs for analyzing 21st-century competencies in postsecondary curricula. We extend prior work to a broader setting that spans multiple types of curriculum documents, academic disciplines, and competency frameworks. Our goal is to assess LLMs’ current capabilities and limitations and to identify strategies for improving LLM-assisted pedagogical reasoning in curricular analytics. Specifically, we address the following research questions:

\begin{enumerate}
    \item What types of curriculum documents are more informative for conducting 21st-century competency analytics?
    \item What is the performance of zero-shot LLMs as a baseline for 21st-century competency analytics tasks, and what are the challenges?
    \item Do reasoning-based prompting strategies address these challenges and improve the reliability of LLM-based 21st-century competency analytics?
\end{enumerate}

The contribution of this study is threefold. First, we evaluate the feasibility of analyzing 21st-century competencies using different types of digital curriculum documents in postsecondary institutions, providing empirical evidence and practical guidance for data-driven competency analytics. Second, we establish baseline performance benchmarks for off-the-shelf LLMs across diverse curriculum document types and varying levels of conceptual specificity in competency frameworks, highlighting systematic limitations in current models. Third, we introduce a reasoning-based prompt strategy that enhances LLM performance in curricular analytics. Together, these contributions advance the empirical and methodological foundations for using generative AI to support large-scale, evidence-based curricular analytics in postsecondary education.

\section{Related Work}
\subsection{Mapping 21st-Century Competencies in Postsecondary Curricula}
Postsecondary institutions are increasingly expected to prepare graduates not only with disciplinary knowledge but also with a broad set of competencies necessary to navigate an increasingly complex and dynamic workforce ~\citep{oecd2023}. These competencies, often referred to as \textit{21st-century competencies}, \textit{21st-century skills}, or \textit{future skills}, are not singular task-specific skills but a diverse, interdisciplinary set of transferable capabilities ~\citep{griffin2012assessment}. Over the past decade, the 21st-century competencies have been a dynamic and rapidly evolving concept,  as various governmental, academic, non-profit, and corporate entities have accelerated the revision of regional frameworks to better respond to evolving economic development, societal priorities, and technological advancements. A systematic review from Cambridge ~\citep{kotsiou2022scoping} identified 99 21st-century competency frameworks developed since 2010, encompassing 341 distinct terms and constructs. Although there is no single shared definition, several competencies commonly found across frameworks include creativity and innovation, critical thinking, problem-solving, communication and collaboration, and information and ICT literacy, among others. 

Despite the growing institutional commitment, the systematic integration of 21st-century competencies into curricula remains relatively limited and uneven ~\citep{oecd2023,oecd2023digital,buckingham2016learning}. The absence of a shared definition and understanding of how such competencies develop across disciplines and contexts impedes curriculum reform. Practical challenges further constrain progress, such as difficulties in assessing 21st-century competencies, limited faculty training and pedagogical support, resource constraints that restrict the development of active learning environments, and rigid program structures that limit curricular innovation and skill-oriented instructional approaches. Therefore, while the value is widely recognized, more systematic and scalable approaches are needed to assess how effectively these 21st-century competencies are embedded within and across curricula and to generate evidence to inform continuous curriculum improvement and institutional decision-making.

In this context, curricular analytics is a promising approach to address these challenges. As a subfield of Learning Analytics (LA), curricular analytics leverages computational approaches to provide evidence-based insights to drive curriculum decision-making and program improvement ~\citep{dawson2014curriculum, greer2016learning}. Over the past few decades, a growing body of curricular analytics research has leveraged curriculum documents such as ~\citep{light2024student}, textbooks ~\citep{li2024,yang2023analyzing}, general catalogs ~\citep{jovanovic2025curriculum,ohland2002creating,irwin2002characterizing,russell2024mapping}, and syllabi ~\citep{zamecnik2024mapping,fiesler2020we,gorski2009we,hong2009understanding,homa2013analysis} , learning management system records ~\citep{shorman2024curriculum} to explore various aspects of curriculum design and delivery topics, such as learning objectives and workforce alignment ~\citep{doyle2025,tan2023,lohr2025leveraging,nguyen2024rethinking,zhang2022skillspan,nguyen2024rethinking,zamecnik2024mapping,javadian2024course, javadian2024course,zamecnik2024mapping,russell2024mapping}, pedagogical strategies ~\citep{liu2024hita,kozov2024,lyu2024evaluating}, course delivery ~\citep{tang2016data}, and assessment methods ~\citep{jovanovic2025curriculum}.

Although skill coverage has long been a well-established topic in curricular analytics, prior research has primarily focused on domain- and task-specific skills. For example, ~\citep{kawintiranon2016} analyzes the coverage of 13 computer science-specific knowledge areas in the Computer Engineering Curricular Guideline (CE2016) ~\citep{durant2015ce2016} from the course descriptions of five universities.  Similarly, studies such as ~\citep{light2024student, russell2024mapping, kitto2020towards}  map course descriptions to job skill profiles derived from Lightcast (formerly Burning Glass Technologies) using open-access general catalogs. Other representative studies, such as ~\citep{javadian2024course}, align curriculum texts with detailed work activities from O*NET, a U.S. Department of Labor database containing thousands of real-world task descriptions across occupations.  Recent studies have explored interdisciplinary and higher-order competencies in curricula using LLMs, as these competencies require contextual inference beyond surface-level semantic matching. ~\citep{jovanovic2025curriculum,jayalath2025scaling,zamecnik2024mapping} analyzes the enterprise  skills and graduate qualities such as self-management, problem-solving,  ethical awareness, oral communication, international perspective, and so on. These early efforts highlight both the potential and key challenges in assessing 21st-century competencies: results depend heavily on the quality and informativeness of curriculum documents, model performance varies across skills ~\citep{zamecnik2024mapping}, and LLMs often struggle with competencies that are implicit or require deep pedagogical interpretation. Building on the gaps identified in previous single-framework and single-domain studies, these findings underscore the need for broader, cross-domain investigations that integrate multiple data types, disciplines, and competency models.

\subsection{Generative AI for Curricular Analytics}

The analysis of 21st-century competencies is closely tied to advances in text mining techniques. Early approaches relied on rule-based methods or keyword matching, which struggled to capture complex constructs. The introduction of transformer-based models marked a significant shift, enabling semantic analysis through contextual embeddings and improving performance for well-defined, domain-specific skills. However, these methods remain limited when addressing abstract competencies like those in 21st-century frameworks, which require not only semantic alignment but also pedagogical inference from course materials such as learning activities and assessments. 

The advent of GenAI has expanded these possibilities. Among various technical strategies, zero-shot prompting with carefully designed task guidance prompts has been examined to evaluate the baseline capability of LLMs in performing competency analytics tasks. For instance, ~\citep{sridhar2023harnessing} and ~\citep{jayalath2025scaling} evaluated GPT-4 in zero-shot settings for tasks such as generating learning objectives and mapping seven enterprise skills to assessments. They found that while performance varied across different skills, generative AI achieved an overall alignment comparable to expert annotations. However, some other studies report more limited performance. For example, ~\citep{zamecnik2024mapping} compared zero-shot and retrieval-augmented generation (RAG) methods for extracting seven graduate attributes from curriculum documents, while ~\citep{xu2025course} extended this by evaluating multiple LLMs across four types of curriculum documents and comparing the performance against traditional NLP approaches.  Their findings show that although RAG improves accuracy for well-defined tasks, LLMs in zero-shot mode alone remain unreliable for consistent skill-level analysis. Parallel efforts in the labor market domain, using GenAI to extract skills from job postings and resumes ~\citep{nguyen2024rethinking}, illustrate faster progress due to the availability of standardized taxonomies and large-scale annotated corpora ~\citep{decorte2022design}. This has enabled experimentation beyond prompt-based methods ~\citep{nguyen2024rethinking,senger2024deep}, including supervised learning and fine-tuning approaches ~\citep{thakrar2025enhancing,herandi2024skill}. Recent studies have also leveraged synthetic data generation and data augmentation to improve the recognition of rare or implicit skills ~\citep{decorte2022design,senger2024deep,myronenko2024improving}.

In summary, while GenAI-based approaches are expanding the analytical capacity of curricular analytics, current practices remain largely prompt-based due to the scarcity of benchmark datasets for postsecondary education. The key technical gap lies in the current LLMs’ limited ability to perform in-context reasoning about how competencies are developed and demonstrated within curricular documents. The challenge stems from both the abstract nature of 21st-century competencies and the heterogeneous structure and quality of curriculum documents ~\citep{nguyen2024rethinking,herandi2024skill}. So technically, this makes higher-order competency analytics fundamentally different from prior task-specific skill mapping, shifting the problem space from information retrieval and classification toward contextual inference and pedagogical reasoning.

\subsection{Reasoning-Based Prompt Engineering}
Among various prompting strategies, a line of research focuses on enhancing LLM reasoning through prompt engineering, which is directly relevant to our analytical scenario. This line of work began with the chain-of-thought (CoT) approach, inspired by the “scratchpad” idea in ~\citep{nye2021show}, in which models improve on multi-step tasks by showing intermediate steps. Building on this idea, Wei et al. ~\citep{wei2022chain} introduced standard CoT prompting, in which few-shot examples guide the model to output step-by-step reasoning, followed by the final answer. This significantly improved performance on math problems, such as those in the GSM8K benchmark. Later, ~\citep{kojima2022large} showed that simply adding “Let’s think step by step” enabled zero-shot CoT, achieving performance close to few-shot versions. To reduce reliance on human-written examples, Zhang et al. ~\citep{zhang2022automatic} proposed Auto-CoT, using an LLM to generate its own reasoning traces. Shao et al ~\citep{shao2023synthetic} took this further by generating new examples via a backward question generation and forward answer synthesis process. Researchers also found that the structure of the reasoning chain matters. Least-to-Most Prompting ~\citep{zhou2022least} breaks down complex problems into simpler sub-problems, improving results over standard CoT. Tree-of-Thought prompting ~\citep{yao2023tree} explores multiple reasoning paths like a search tree, helping with planning and strategy tasks. 

More recently, prompting approaches that explicitly decompose inputs into key semantic components have emerged. For example, Summary CoT (SumCoT) ~\citep{wang2023element} encourages models to summarize key contextual elements (e.g., who, what, when, why) before generating final responses, while Rephrase-and-Respond (RaR) ~\citep{deng2024rephrase} guides models to restate questions in semantically distinct yet complementary ways before answering, thereby enhancing reasoning diversity and robustness. These methods share a common intuition: structuring model reasoning around core semantic components can support a more systematic interpretation of complex inputs. This intuition is particularly relevant in curricular analytics. From a learning sciences and pedagogical perspective, 21st-century competencies are developed through instructional strategies and learning activities that are intentionally designed and context dependent ~\citep{retnawati2018teachers,ghanizadeh2020higher}. As a result, meaningful pedagogical inference depends on identifying and interpreting the instructional components that encode these learning intentions. This perspective aligns with curricular analytics tasks, in which valid reasoning requires recognizing and synthesizing key pedagogical information embedded in course materials.

\section{Study Context}
\begin{figure*}[h]  
    \centering
    \includegraphics[width=1\textwidth]{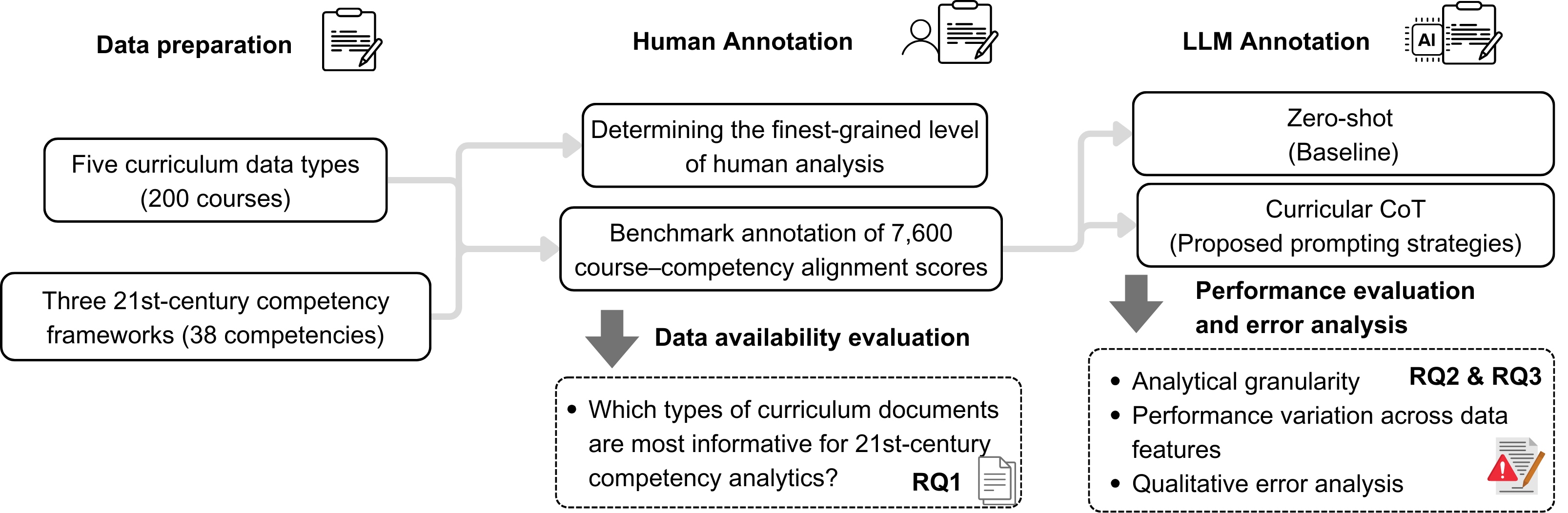}
    \caption{Overall study design}
    \label{fig:pipeline}
\end{figure*}

Figure~\ref{fig:pipeline} shows the overall design of this study. For \textit{RQ1}, we identify representative types of curriculum documents and conduct a rigorous human annotation process to assess their value for 21st-century competency analytics. This process yields a benchmark dataset of 7,600 pairs of curriculum-competency alignment scores. For \textit{RQ2}, we evaluate the performance of zero-shot LLMs as a baseline on this benchmark through quantitative and qualitative analyses, focusing on limitations in analytical granularity and sensitivity to curriculum and competency characteristics. Finally, for \textit{RQ3}, we assess the effectiveness of proposed reasoning-based prompting strategies in improving LLMs’ competency analytics performance by addressing the identified limitations. 

\subsection{Data}
\subsubsection{Curriculum documents}
We draw on ~\citep{de2024curriculum}, which provides a systematic review of data used in curricular analytics, to identify representative types of digital curriculum documents commonly available at postsecondary institutions. Table~\ref{tab:datatypes} summarizes five types of curriculum documents, categorized by pedagogical element coverage and data source.
\begin{table}[H]
    \centering
    \small
    \begin{tabular}{p{3.5cm} p{3cm} p{9cm}} 
    \toprule
   Data type&  Data source& Potential pedagogical element coverage\\ 
        \midrule
          Concise course description&  General catalog& High-level course topics; learning goals or competencies; general instructional format; high-level assessment overview\\ 
                Detailed course description&  Course syllabus& Course topics; articulated learning goals; planned learning activities; instructional or delivery format; assessment plan\\
  Learning objective&  Syllabi& Explicit learning goals, such as knowledge domains, skills, and competencies\\
  Instructional schedule& Syllabi&Temporal organization of topics; sequencing of instructional activities; specified learning activity types; assessment plan\\
               Learning activity content&  Learning management system record& Detailed learning tasks and prompts; instructional materials; learning activity formats; assessment structure and grading components\\
\bottomrule
\end{tabular}
\caption{Overview of curriculum document types for curricular analytics in postsecondary education}\label{tab:datatypes} 
\end{table}

We then create samples of these curriculum document types from three distinct sources:

\begin{itemize}
    \item  \textit{Course syllabi} from Open Syllabus\footnote{\url{https://www.opensyllabus.org/}}, a nonprofit organization maintaining an archive of over 20.9 million course syllabi from postsecondary education institutions worldwide.
    \item \textit{General catalog} from publicly accessible websites of two institutions in the United States: a large, urban, public two-year college and a public four-year university.
    \item \textit{Learning management system} data from a public four-year university in the United States, including instructor-generated prompts and instructions for various learning activities as logged within the Canvas learning management system.
\end{itemize}

From the general catalog, we extract short-form descriptions and basic metadata of individual courses, including the course title and department. From course syllabi, we extract overall descriptions, learning goals, and activity schedules for each course. From learning management system data, we retrieve full instructional prompts and content for learning activities (e.g., assignments, discussions, and quizzes) and concatenate them at the course level. For comparability across data types, we apply stratified sampling by subject area and document length for each data type, followed by a manual review to remove low-quality or incomplete samples. This process yields five datasets, each containing 40 course-level samples, for a total of 200 curriculum documents.

\subsubsection{Competency frameworks}
We select 21st-century competency frameworks that were: (1) published after 2010, (2) issued by authoritative organizations (e.g., government agencies or international institutions), (3) applicable to postsecondary education, and (4) characterizing general, cross-disciplinary competencies rather than domain-specific skills. Table~\ref{tab:frameworks} summarizes the three frameworks selected for this study, which vary in definitional granularity. More detailed information on these frameworks can be found in the Appendix Tables~\ref{tab:Summary of competency items},~\ref{O*NET(Workforce Competencies)},~\ref{EU(Key competences)}, and~\ref{ESDC (Success model)}.

\begin{table}[ht]
    \centering
    \begin{tabular}{p{5.5cm} p{2cm} p{1cm} p{3.5cm} p{2.5cm}} 
    \toprule
  Competency frameworks& Country& Year& Education / Labor market& Count of competencies\\ 
        \midrule
         O*NET (Workforce Competencies) \tablefootnote{\url{https://files.eric.ed.gov/fulltext/EJ1109948.pdf}}& US&2013& Labor market&  21\\ 
               EU (Key Competences)\tablefootnote{\url{https://op.europa.eu/en/publication-detail/-/publication/297a33c8-a1f3-11e9-9d01-01aa75ed71a1/language-en}}& European Union&2018& Education&  8\\
              ESDC (Success Model)\tablefootnote{\url{https://www.srdc.org/project/Research-report-to-support-the-launch-of-Skills-for-Success-Structure-evidence-and-recommendations-Final-report/}}& Canada& 2024& Labor market&   9\\
\bottomrule
\end{tabular}
\caption{Overview of 21st-century competencies frameworks included in this study}\label{tab:frameworks} 
\end{table}

\subsection{Human Annotation}

\subsubsection{Rubrics for curriculum-competency mapping}
We recruited two graduate students in education with expertise in curriculum and instructional design. During the tutorial session, the annotators were introduced to the task, and each received three sample curriculum documents from each of the five curriculum document types (15 in total), which is a holdout set to avoid circularity between rubric development and model evaluation. They independently annotated the samples and refined the author’s draft rubric, adjusting its levels and descriptions to enhance the alignment between competencies and course documents. Finally, the author and two annotators jointly finalized a version that defines the finest level of analytical granularity achievable by human raters across all data and framework types, as shown in Table~\ref{tab:rubric}.

\begin{table}[H]
    \centering
    \begin{tabular}{p{1.2cm} p{14cm} } 
    \toprule
  Score & Description \\ 
        \midrule
         3 & The competency is explicitly and clearly stated as a course objective.  \\
 2 &The competency can be reasonably inferred from the course document.  \\
 1 &The competency is vaguely implied but not clearly emphasized.  \\ 
               0 & The competency is clearly unrelated to the course..  \\
              NA& The competency may be relevant, but there is insufficient information to determine its presence.  \\           
\bottomrule
\end{tabular}
\caption{Curriculum-competency alignment rubric}\label{tab:rubric} 
\end{table}
\subsubsection{Annotation and inter-rater reliability}
Before the formal annotation, we conduct a calibration round in which two annotators annotate 15 curriculum documents from each of the five document types (45 samples) for each of the three competency frameworks (38 competencies), which is 1,710 curriculum-competency pairs in total. The two annotators achieve a Cohen's  $\kappa$ of 0.841 in the framework of EU (Key Competences), a $\kappa$ of 0.288 in the framework of O*NET (Workforce Competencies), and a $\kappa$ of 0.168 in the framework of ESDC (Success Model). Follow-up discussions reveal that most disagreements stem from differing interpretations of whether course context (e.g., subject matter and course type) should inform competency inference, leading to confusion between codes \textit{0} (clearly irrelevant) and \textit{NA} (insufficient information). We therefore refine the coding guidelines to explicitly incorporate course context: competencies clearly outside the instructional scope were coded as \textit{0}, whereas potentially relevant competencies lacking sufficient textual evidence were coded as \textit{NA}. After aligning this understanding, the two annotators re-annotate all samples in the calibration round. The final inter-rater reliability on these two frameworks is $\kappa$ = 0.94 in O*NET (Workforce Competencies), and $\kappa$ = 0.92 in ESDC (Success Model). 

\begin{figure}
    \centering
    \includegraphics[width=1\linewidth]{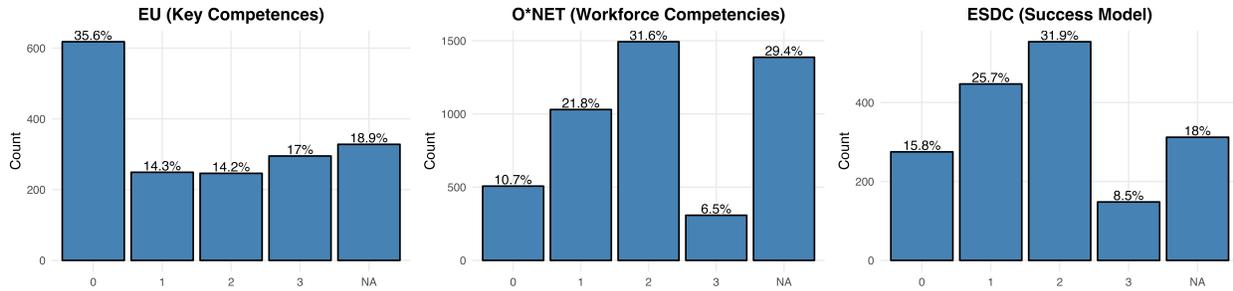}
    \caption{Distribution of human annotation scores across competency frameworks}
    \label{fig:annotation}
\end{figure}

Then, the two annotators independently work on the remaining dataset, scoring each competency in each framework for each document. To maintain annotation reliability during this phase, random sampling and periodic cross-checks are implemented. Specifically, each annotator periodically reviews a small subset of the other’s annotations (approximately 10\%–15\%), and any discrepancies are resolved through discussion. This annotation process yields a benchmark dataset comprising 200 course documents annotated across the three frameworks (38 competencies in total), resulting in 7,600 pairs of curriculum-competency alignment scores. Figure~\ref{fig:annotation} presents the distribution of human annotations, showing the count and percentage of each rubric score across the five data types. A detailed summary of the annotation distribution for each data type is provided in Table~\ref{annotation_distribution} in the Appendix.

\subsection{LLM Annotation}

\subsubsection{Models}
We evaluate LLM performance on the 7,600 human-annotated curriculum-competency alignment pairs using four models: GPT 3.5-turbo and GPT-4o (OpenAI, proprietary), and Llama-3-70B and Llama-3-8B (Meta AI, open-weight). In all experiments, the generation temperature was set to 0.

\subsubsection{Prompting strategies}
\begin{figure}[htbp]
    \centering
    \includegraphics[width=1\textwidth]{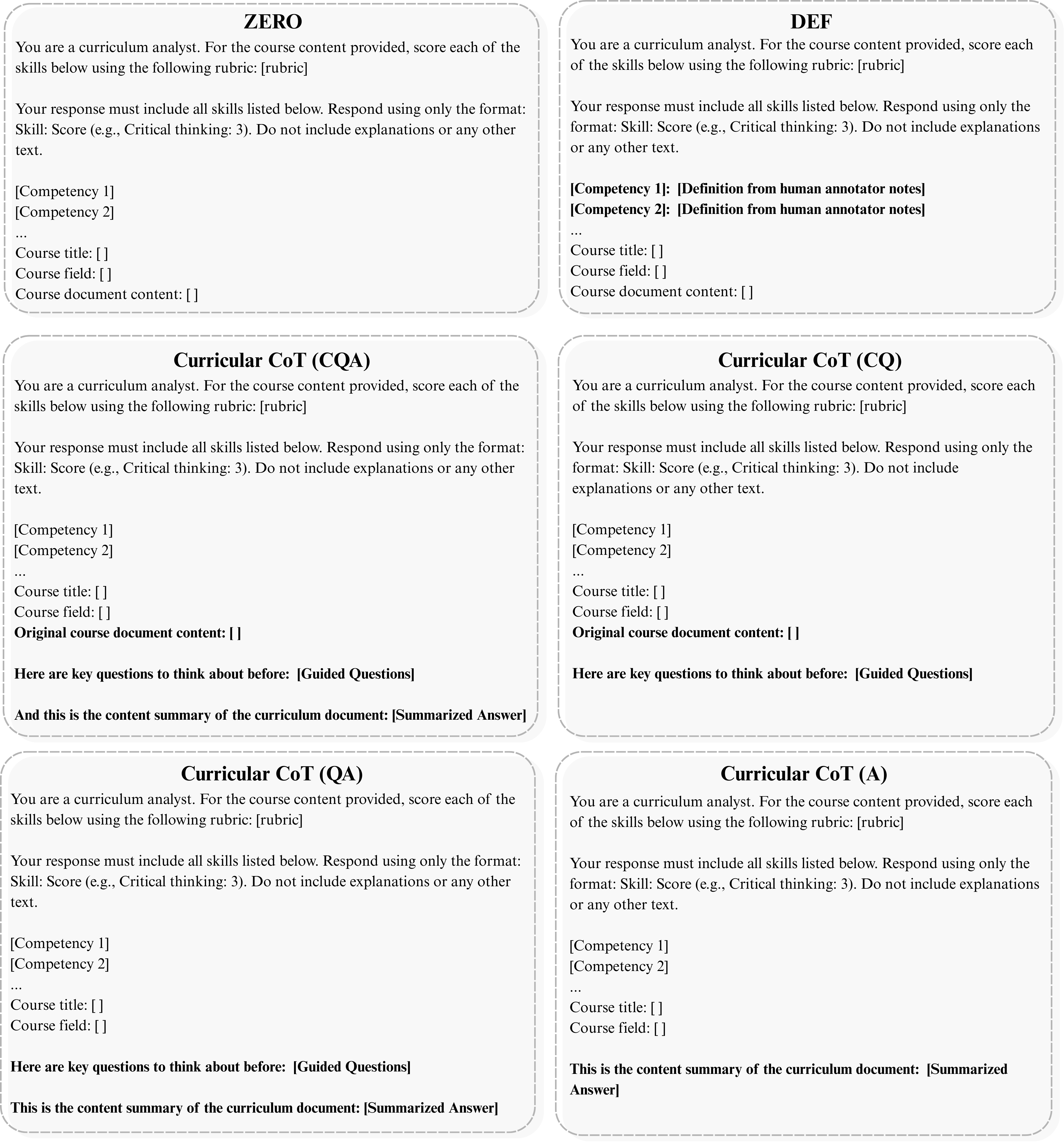}
    \caption{Detailed templates for each prompting strategy}
    \label{fig:prompt}
\end{figure}

We examine six prompting strategies: a zero-shot baseline, zero-shot prompting with competency definitions, and four variants of our proposed reasoning-based approach. Full prompt templates are provided in Figure~\ref{fig:prompt}.
\begin{enumerate}
    \item \textbf{Zero-Shot (ZERO):} The LLM is provided with the course document content and a list of competency names. It reflects the model’s default reasoning ability without any additional scaffolding or contextual support. 
    \item \textbf{zero-shot prompting with competency definitions(DEF):} The model is provided with the course document, competency names, and corresponding definitions from human annotator notes. This strategy aims to reduce competency ambiguity and improve interpretive consistency by grounding model reasoning in explicit human-defined criteria.
    \item \textbf{Curricular CoT:} Building on prior work ~\citep{wang2023element,deng2024rephrase}, we introduce a structured reasoning-based prompting strategy for curricular analysis. The model follows a two-step reasoning process:
    
\begin{itemize}
    \item \textit{Element extraction:} Extracting key pedagogical components using guided questions (Table~\ref{tab:guided questions}) grounded in curriculum design theory.
    \item \textit{Standardized content representation:} The extracted elements are synthesized into a structured summary that reorganizes unstructured curriculum content into a consistent representation, which is then used in the final competency evaluation prompt.
\end{itemize}

    This design mitigates traditional challenges in curricular analysis arising from heterogeneous document granularity ~\citep{arafeh2016curriculum}, inconsistent wording and structure ~\citep{tian2024enhancing}, and long or variable input length. We further distinguish the following variants of this prompting strategy.

\begin{table}[H]
    \centering
    \begin{tabular}{p{5.5cm} p{5cm} p{6cm}} 
    \toprule
    Course Description & Learning Objectives & Learning Activities / Instructional Schedule \\ 
    \midrule

1.\textbf{Course focus:} What is the primary content focus of the course?  

2.\textbf{Core knowledge and skills:} What key knowledge and skills are emphasized in the course?  

3.\textbf{Primary learning tasks:} What major tasks are students expected to complete?  

4.\textbf{Instructional format:} How is the course delivered (e.g., lecture, discussion, project-based, hybrid)?  

5.\textbf{Assessment approach:} How is student learning evaluated in the course?  

&

1.\textbf{Target knowledge:} What domain-specific knowledge are students expected to acquire?  

2.\textbf{Target skills:} What skills are students expected to develop?  

3.\textbf{Expected performance:} What tasks or capabilities should students be able to demonstrate upon course completion?  

&   

1. \textbf{Activity summary:} What is the learning activity (one sentence)?  

2. \textbf{Activity type:} What type of activity is it (e.g., discussion, assignment, project, lab)?  

3. \textbf{Target knowledge and skill:} What knowledge and skill do the activity address?  

4. \textbf{Student deliverable:} What is the expected student output for this activity?  

5. \textbf{Assessment method:} How is performance on this activity evaluated?  

\\
    \bottomrule
    \end{tabular}
    \caption{Guided questions by curriculum document type}
    \label{tab:guided questions} 
\end{table}

\begin{enumerate}
\item \textbf{Curriculum + Questions + Answers (CQA):} The LLM receives the original curriculum document, competency names, guided questions, and the corresponding summarized answers, and is instructed to reflect on this information before analyzing competency alignment.

\item \textbf{Curriculum + Questions (CQ):} The LLM receives the original curriculum document, competency names, and guided questions, and reflects on the questions before analyzing competency alignment.

\item \textbf{Questions + Answers (QA):} The LLM receives the summarized curriculum content, competency names, and guided questions, and reflects on the questions and summarized content before analyzing competency alignment.

\item \textbf{Answers Only (A):} The LLM receives only the summarized curriculum content and competency names to analyze competency alignment.
\end{enumerate}
\end{enumerate}

\begin{figure}[ht]
    \centering
    \includegraphics[width=0.8\textwidth]{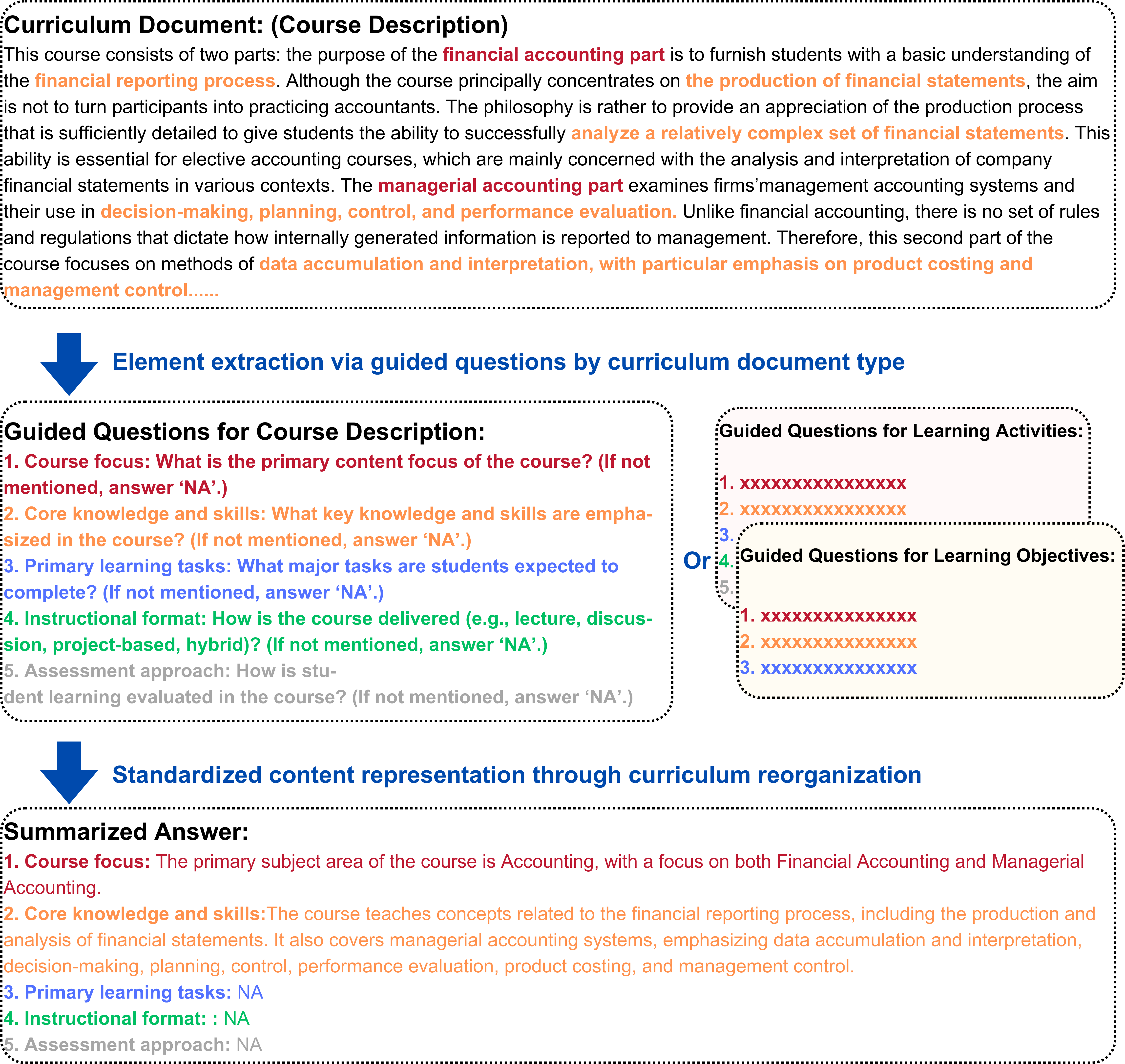}
    \caption{Full pipeline and example of \textbf{Curricular CoT}}
    \label{fig:cot}
\end{figure}

\subsection{Performance Evaluation}

\subsubsection{Evaluation metrics}
In this study, model performance is evaluated using Accuracy and macro-averaged Precision, Recall, and F1 score. Macro-averaged metrics are reported to account for class imbalance in the ground-truth data, as shown in Figure~\ref{fig:annotation} and detailed in Table~\ref{annotation_distribution} in the Appendix. To capture the ordinal structure of the labels and assess agreement with human annotators, we additionally report the Weighted Cohen’s Kappa and the Intraclass Correlation Coefficient (ICC).

\subsubsection{Analytical granularity}
We further examine LLMs’ ability to distinguish fine-grained representations of 21st-century competencies in curriculum documents. Table~\ref{classification levels} summarizes the hierarchical taxonomy used for performance evaluation, in which detailed competency categories are progressively merged into coarser levels. We first evaluate performance at the finest granularity that human annotators could reliably achieve: a five-class classification task capturing nuanced skill distinctions. We then gradually reduce the level of granularity to four-, three-, and two-class settings to assess how performance changes as task complexity decreases. We adopt an aggregation-based evaluation approach, collapsing predictions from the 5-level taxonomy into coarser levels before computing performance metrics. 

\begin{table}[H]
\centering
\small
\renewcommand{\arraystretch}{1.3}
\setlength{\tabcolsep}{5pt}

\begin{tabular}{|p{4cm}|p{3.7cm}|p{3.7cm}|p{4cm}|}
\hline
\textbf{5-Level} & \textbf{4-Level} & \textbf{3-Level} & \textbf{2-Level} \\
\hline
3: Explicitly stated&
3: Explicitly stated&
3: Explicitly stated&
\multirow{3}{*}{1 \& 2 \& 3: Broadly Covered} \\
\cline{1-3}

2: Reasonably inferred&
2: Reasonably inferred&
\multirow{2}{*}{%
  \begin{tabular}[t]{@{}l@{}}
  1 \& 2: Partially inferred
  \end{tabular}
} & 
\\
\cline{1-2}

1: Possibly implied&
1: Possibly implied&
 & 
\\
\hline

0: Unrelated&
\multirow{2}{*}{%
  \begin{tabular}[t]{@{}l@{}}
  0 \& NA: Unrelated or\\ Insufficient information
  \end{tabular}
} &
\multirow{2}{*}{%
  \begin{tabular}[t]{@{}l@{}}
  0 \& NA: Unrelated or\\ Insufficient information
  \end{tabular}
} &
\multirow{2}{*}{%
  \begin{tabular}[t]{@{}l@{}}
  0 \& NA: Unrelated or\\ Insufficient information
  \end{tabular}
} \\
\cline{1-1}

NA: Insufficient information&
 & 
 & 
\\
\hline
\end{tabular}
\caption{Description of granularity levels used for LLM classification evaluation}
\label{classification levels}
\end{table}

\subsubsection{Performance heterogeneity across data features}  
We run a series of regression analyses to examine how LLM performance varies across curriculum document type, model version, and competency framework, and to identify data characteristics associated with greater difficulty in competency analysis. The dependent variable $Y_i$ is specified in two forms: (1) \textit{Prediction Accuracy}, a binary indicator denoting whether the LLM prediction exactly matches the human annotation, and (2) \textit{Score Difference}, a continuous measure capturing the deviation between LLM-predicted and human-annotated competency scores. The regression model is specified as:

\begin{equation}
Y_i = \alpha 
+ \beta\cdot \mathbf{DataType}_i
+ \gamma \, \text{WordCount}_i
+ \delta\cdot \mathbf{Model}_i
+ \sigma \cdot \mathbf{Framework}_i
+ \eta \cdot \mathbf{SubjectMatter}_i
+ \varepsilon_i
\label{reg:1}
\end{equation}
where $\mathbf{DataType}_i$, $\mathbf{Model}_i$, and $\mathbf{Framework}_i$ are categorical indicators. Learning Objectives (data type), GPT-4o (LLM model), and the EU Key Competences framework serve as reference categories, as they exhibit the highest overall performance. $\text{WordCount}_i$ controls for variation in text length. Subject matter is included as a set of categorical controls, STEM, Humanities and Social Sciences, Applied Disciplines, and Other/Interdisciplinary, to account for disciplinary context. We estimated separate regressions for each classification granularity (from 2-class to 5-class) and for each dependent variable (accuracy and score difference).

\section{Results}
\subsection{What types of curriculum documents are more informative for conducting 21st-century competency analytics? (RQ1)}
\begin{figure*}[ht]  
    \centering
    \includegraphics[width=0.8\textwidth]{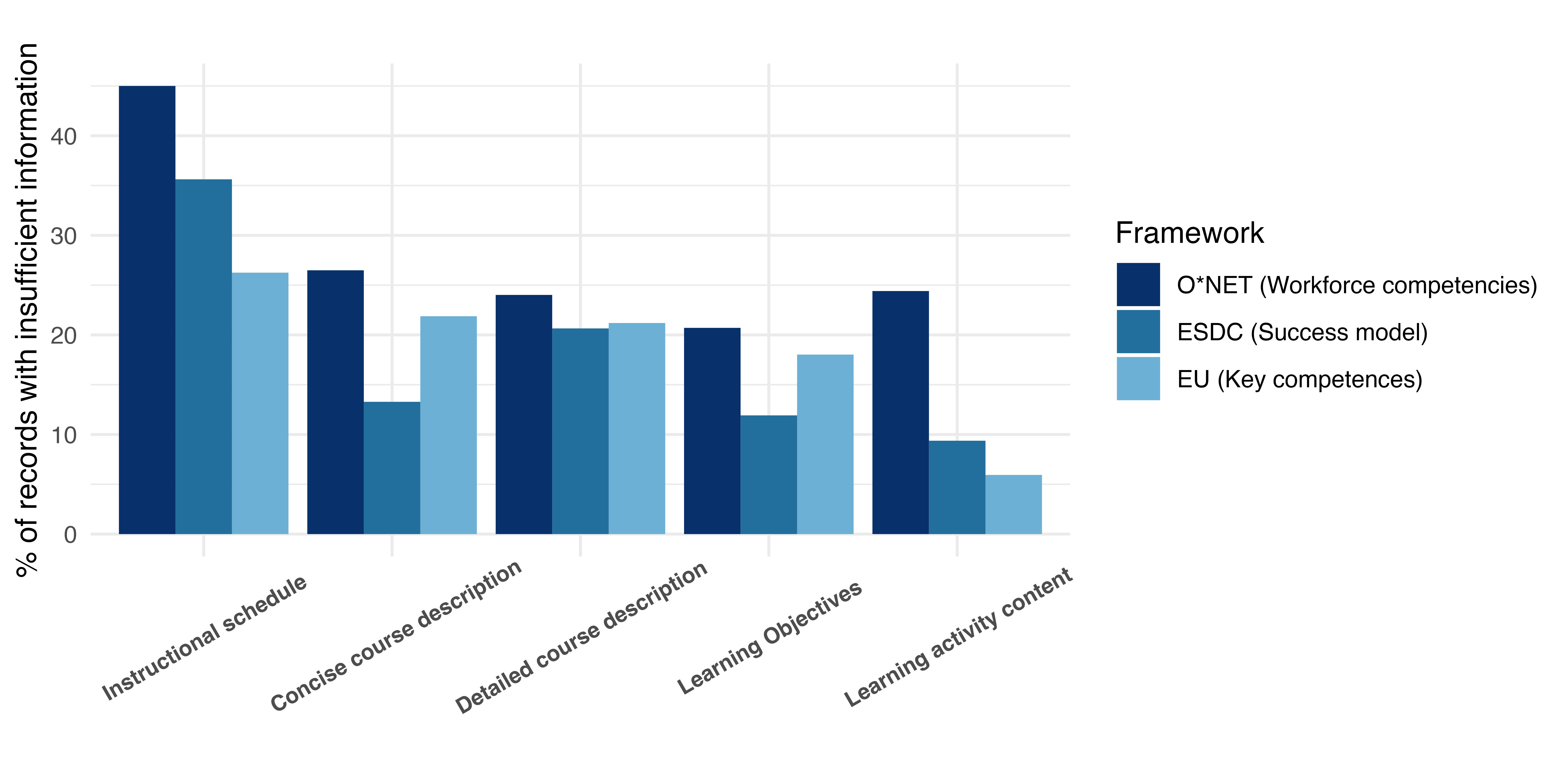}
    \caption{Percentage of curriculum documents with insufficient information (NA: competency may be relevant, but available content is insufficient to determine its presence) across competency frameworks and curriculum document types}
    \label{fig:percentage-na}
\end{figure*}

During human annotation, annotators could assign an NA label when curriculum documents lacked sufficient information to determine whether a competency was covered. We use the proportion of NA annotations as an indicator of each document type’s usability for 21st-century competency analytics. Figure~\ref{fig:percentage-na} shows the percentage of documents labeled as NA based on finalized annotations after inter-rater reliability was established.

Across all three competency frameworks, instructional schedules exhibit the highest rates of insufficient information for competency inference—up to 45\% under ONET (Workforce Competencies), 36\% under ESDC (Success Model), and 27\% under the EU (Key Competences). In contrast, learning objectives and learning activity content provide clearer pedagogical cues and show the lowest NA rates. Usability also varies by framework granularity: finer-grained frameworks, such as ONET’s distinction between “written expression” and “written comprehension,” are associated with higher NA rates, while broader competency definitions (e.g., ESDC’s general “writing” category) are more readily inferred from limited information.

Overall, detailed descriptions of specific learning activities provide the richest information for evaluating 21st-century competencies. Learning objectives and course descriptions can also provide helpful cues, whereas instructional schedules generally lack the depth needed to assess 21st-century competencies. However, we note that these findings are specific to our sample and may not reflect general patterns. Information sufficiency ultimately depends on both the data type and the targeted competency, and it should be determined in the context of the specific datasets.

\subsection{What is the performance of zero-shot LLMs as a baseline for 21st-century competency analytics tasks, and what are the challenges? (RQ2)}

After assessing the usability of the five curriculum document types for 21st-century competency analytics, we evaluate LLMs’ baseline performance in a zero-shot setting, reflecting their default reasoning ability without additional scaffolding or contextual support.

\subsubsection{Performance variation across analytical granularity}
\begin{figure*}[ht]  
    \centering
    \includegraphics[width=0.8\textwidth]{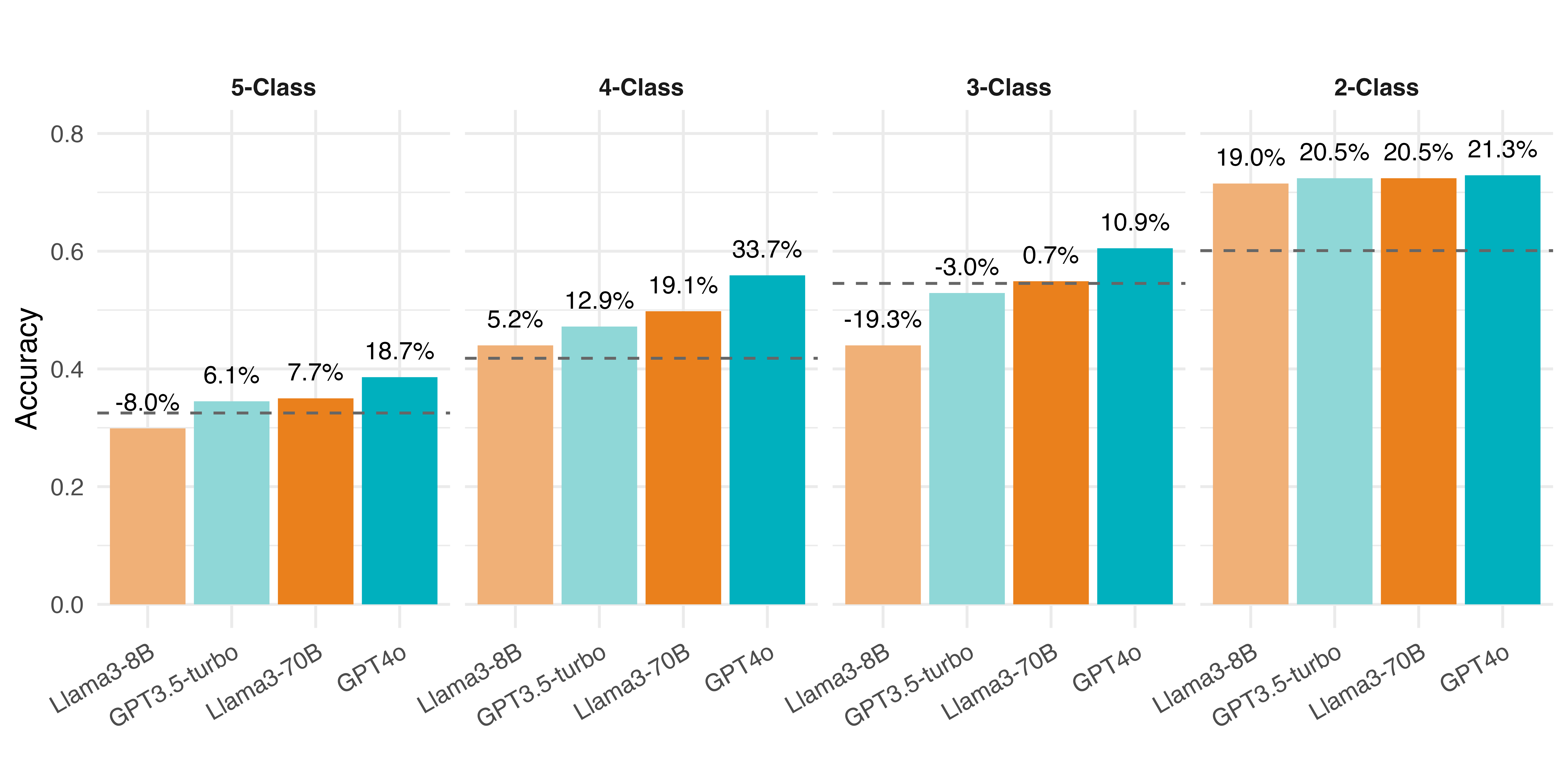}
    \caption{Zero-shot accuracy across levels of classification granularity and comparison with random guessing}
    \label{fig:rq2}
        \begin{tablenotes}
\footnotesize
\item \textbf{Note:} Accuracy here represents averages across data types within each competency framework. Instructional schedules are excluded due to insufficient information in many records. The gray dashed line marks the Majority Class Guessing (MCG) baseline, which accounts for class imbalance, defined as $\text{Expected Accuracy}=\max_i P(i)$. MCG was computed for each data type within each framework, and the final score reflects their average. Numbers on the bars indicate the deviation from the corresponding MCG accuracy.
\end{tablenotes}
\end{figure*}

We first examine how LLM baseline performance varies with the granularity of competency classification. As shown in Figure~\ref{fig:rq2}, more advanced models such as GPT-4o and Llama3-70B consistently achieve higher prediction accuracy across all levels of classification granularity, particularly in more complex, finer-grained tasks. In the three- to five-class settings, smaller models such as Llama3-8B often perform at or near the random-guessing baseline, whereas GPT-4o exceeds the baseline by 10.9\%–33.7\%. This gap indicates that greater model scale and reasoning capacity are critical for handling nuanced competency distinctions in zero-shot settings.

Despite these differences, LLMs overall exhibit limited effectiveness in fine-grained competency analysis compared to humans. In the five-class setting, all models perform only marginally above random, suggesting a shared difficulty in distinguishing among different levels of competency coverage. Performance improves substantially as the task becomes easier. When reduced to a binary classification task, as shown in Table~\ref{classification levels}, all models achieve accuracies above 70\% (GPT-3.5-turbo: 72.4\%; GPT-4o: 72.9\%; Llama3-70B: 71.5\%; Llama3-8B: 72.4\%). Full results for additional evaluation metrics are reported in Table~\ref{tab:rq2 All performance metrics} in the Appendix.

\subsubsection{Performance heterogeneity across data features}
We next examine which data characteristics pose greater challenges for LLMs in producing accurate and reliable predictions. Table~\ref{tab:1} reports the regression results. Across all analytic settings, both curriculum document type and competency framework are significantly associated with variation in LLM prediction accuracy. 

Among curriculum document types, learning objectives (the reference category) consistently show the highest accuracy after controlling for task context. In contrast, instructional schedules are associated with the largest performance decline, with accuracy reductions ranging from 0.114 in the five-class setting to 0.216 in the binary classification task. This is followed by detailed course descriptions, concise course descriptions, and learning activity content from the learning management system. Importantly, this ordering of performance degradation is stable across all levels of classification granularity. 

Framework characteristics also play a substantial role. The EU Key Competences framework, representing the most general level of competency definition, is associated with significantly higher prediction accuracy than the two more fine-grained frameworks across all classification settings. Text length, as measured by word count, is not significantly associated with prediction accuracy. Finally, model capacity matters primarily for fine-grained tasks: more advanced models (e.g., GPT-4o and Llama3-70B) significantly outperform smaller models in multi-class settings, whereas in the binary classification task, accuracy differences across models are no longer statistically significant.

\begin{table}[h]
    \centering
    \scriptsize
    \begin{threeparttable}
    \begin{tabular}{llcccccccc}
    \toprule
          && \multicolumn{4}{c}{Prediction accuracy}& \multicolumn{4}{c}{Score difference}\\
  && 5 Class& 4 Class& 3 Class& 2 Class& 5 Class& 4 Class& 3 Class&2 Class\\ 
        \midrule
         Data type&Concise course description& -0.043***&-0.034***& -0.029***&-0.069***& 0.225***& 0.171***& 0.089***&0.084***\\ 
              &Detailed course description& -0.054***&-0.046***& -0.059***&-0.088***& 0.213***& 0.153***& 0.107***&0.126***\\ 
         &Learning activity content& -0.038**&-0.042**& -0.044***&-0.074***& -0.213***& -0.127***& -0.094***&-0.089***\\
  &Instructional schedule& -0.114***& -0.045***& -0.120***& -0.216***& 0.645***& 0.469***& 0.281***&0.257***\\
  Length &Word count& 0.008·& 0.005& 0.001& 0.023***& 0.030**& 0.031***& 0.015**&0.008·\\
  Model&GPT 3.5-turbo& -0.039***& -0.014*& -0.051***& 0.010& 0.052**& 0.070***& 0.028**&-0.091***\\
  &Llama3-70B& -0.044***& -0.033***& -0.054***& 0.000& 0.132***& 0.128***& 0.082***&-0.029***\\
  &Llama3-8B& -0.020**& 0.038***& 0.000& 0.001& -0.636***& -0.267***& -0.222***&-0.225***\\
 Framework&O*NET& -0.093***& -0.114***& 0.012*& 0.011*& 0.100***& 0.001& 0.053***&-0.003\\
 &ESDC& -0.115***& -0.163***& -0.102***& -0.017**& 0.234***& 0.214***& 0.129***&-0.034***\\ 
\midrule
 &Adjusted $R^2$ & 0.0158&0.018& 0.018&0.027& 0.092&  0.058& 0.061&0.069\\
\bottomrule
    \end{tabular}
    \caption{Regression coefficients for LLM alignment with human annotations across classification granularity.}\label{tab:1} 
    \begin{tablenotes}
\footnotesize
\item \textbf{Note:} Each column presents results from a separate regression conducted at a specific classification granularity. All regressions are run at the course competency level. The dependent variables include: (1) a binary accuracy indicator, which captures whether the LLM exactly predicts the human-annotated score  (1 if the LLM prediction exactly matches the human annotation, 0 otherwise), and (2) a continuous score difference variable (LLM predicted score $-$ human annotated score), which measures the degree of deviation from human judgment, revealing whether it overestimates or underestimates the competency relevance. For each group tested, we use the highest performing condition as Learning Objectives (data type), GPT-4o (model), and EU (Key competences) (framework) as the reference group to assess performance degradation across other conditions. Statistical significance is denoted by asterisks: $p<0.10(\cdot)$, $p<0.05(^*)$, $p<0.01(^{**})$, $p<0.001(^{***})$.
\end{tablenotes}
\end{threeparttable}
\end{table}

For score differences, the regression results indicate that, after controlling for task context, LLMs generally overestimate the extent to which a course covers a given competency. This upward bias is reflected in positive and statistically significant intercepts across all classification granularities (0.128 in the five-class task, 0.019 in the four-class task, 0.051 in the three-class task, and 0.147 in the binary task). Curriculum document characteristics further shape this bias. Data type is a significant predictor of overestimation, with instructional schedules producing the largest positive score differences across all analytic settings, indicating the strongest tendency toward overprediction. Course descriptions also exhibit significant but more moderate overestimation effects. Framework characteristics contribute as well: relative to the most general framework (EU Key Competences), both O*NET (Workforce Competencies) and ESDC (Success Model) are associated with significantly larger score differences. Finally, word count is positively associated with score overestimation, suggesting that longer or more content-rich curriculum documents lead LLMs to infer higher levels of competency coverage.

\subsubsection{Error analysis}\label{4.2.3 error analysis}
\begin{figure*}[htbp]
    \centering
    \includegraphics[width=0.9\textwidth]{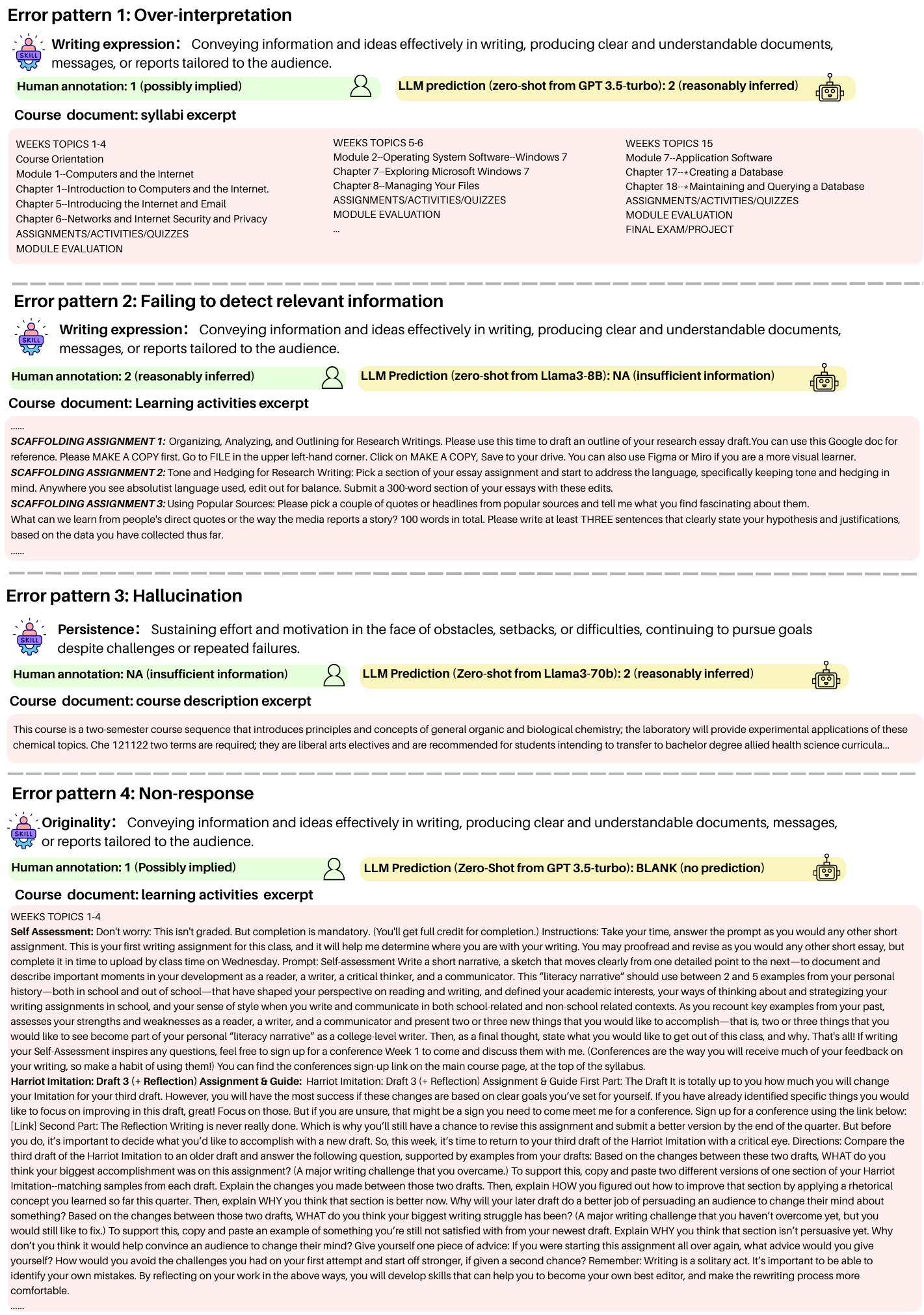}
    \caption{Error patterns in baseline (Zero-Shot) LLM performance for 21st-Century competency analysis}
    \label{fig:error-rq2}
\end{figure*}

We manually analyzed cases where LLM predictions diverged from human annotations and identified several recurring error patterns. Figure~\ref{fig:error-rq2} provides a representative example illustrating these errors.

\begin{enumerate}
   \item \textbf{Over-interpretation:} This error pattern is observed across all models and all curriculum document types we evaluated. A possible explanation is that LLMs tend to overinterpret available content, while human raters are more rigorous in seeking explicit pedagogical evidence. For example, we find that when a course document mentions an exam, human annotators are cautious about inferring writing-related competencies because exam formats can vary widely. In contrast, LLMs tend to default to the assumption that exams are writing-based, leading them to confidently infer writing competencies, which reflects biased reasoning.
    \item \textbf{Failure to detect relevant information:} This error pattern is common in lengthy curriculum document types such as learning activity content, where dense information makes it difficult for LLMs to capture and interpret key details accurately. For example, when information about skill development is embedded within subtle descriptions of teaching activities, human annotators can often detect and reason from these cues, whereas LLMs may overlook such nuanced evidence and incorrectly assign \textit{NA}.
    \item \textbf{Hallucination:} This error pattern is more common in concise course descriptions and instructional schedule data, which contain limited information compared to other data types. In some cases, LLMs make confident predictions even when the text offers little to no evidence to support such inferences.
    \item \textbf{Failure to generate a response}: This error pattern is relatively rare and occurs predominantly in smaller models (e.g., Llama3-8B). In these cases, the model fails to return any prediction. It is most frequently observed in data content types used for learning activities, likely due to their unstructured format or because the input length exceeds the model's context window. 
\end{enumerate}

\subsection{Do reasoning-based prompting strategies address these challenges and improve the reliability of LLM-based 21st-century competency analytics? (RQ3)}

In this section, we evaluate whether our proposed reasoning-based prompting strategies improve LLMs’ pedagogically grounded reasoning for higher-order competency analysis by replicating prior experiments.

\subsubsection{Performance heterogeneity across data feature}
Table~\ref{tab:prompt_accuracy} presents the overall prediction accuracy of five proposed prompting strategies, compared to the baseline zero-shot setting, across all levels of classification granularity. Overall, we find that Curricular CoT  improves prediction accuracy over the zero-shot baseline across all levels of task granularity. However, these improvements are more apparent in larger and more advanced models, such as GPT-4o and Llama3-70B. In contrast, the Definition-based (DEF) strategy, which provides the model with human-annotated notes for each competency, does not improve performance and can even reduce accuracy.

\begin{table}[ht]
\centering
\begin{threeparttable}
\begin{tabular}{llllllllll}
\toprule
      & \multicolumn{4}{c}{GPT3.5-turbo} & & \multicolumn{4}{c}{GPT4o} \\
Prompting & 5 Class & 4 Class & 3 Class & 2 Class & & 5 Class & 4 Class & 3 Class & 2 Class \\
\midrule
\rowcolor{gray!15}
ZERO (Baseline)& 0.337 & 0.457 & 0.536 & 0.712 & & 0.334 & 0.530 & 0.586 & 0.704 \\
DEF       & 0.337 & 0.452 & 0.512 & 0.671 & & \textbf{0.337} & 0.524 & 0.571 & 0.690 \\
CQA       & 0.322 & 0.444 & 0.535 & 0.706 & & \textbf{0.338} & \textbf{0.536} & \textbf{0.599} & \textbf{0.713} \\
CQ        & 0.321 & 0.438 & 0.517 & 0.706 & & 0.334 & \textbf{0.531} & \textbf{0.584} & 0.704 \\
QA        & 0.328 & 0.446 & 0.525 & 0.707 & & \textbf{0.341} & \textbf{0.535} & \textbf{0.594} & \textbf{0.705} \\
A         & 0.332 & 0.454 & 0.533 & 0.712 & & \textbf{0.347} & \textbf{0.538} & \textbf{0.599} & \textbf{0.710} \\
\bottomrule
\end{tabular}
\vspace{4mm}
\begin{tabular}{llllllllll}
\toprule
      & \multicolumn{4}{c}{Llama3-8B} & & \multicolumn{4}{c}{Llama3-70B} \\
Prompting & 5 Class & 4 Class & 3 Class & 2 Class & & 5 Class & 4 Class & 3 Class & 2 Class \\
\midrule
\rowcolor{gray!15}
ZERO (Baseline)& 0.315 & 0.442 & 0.442 & 0.710 & & 0.334 & 0.473 & 0.535 & 0.701 \\
DEF       & 0.305 & \textbf{0.455 }& \textbf{0.455} & 0.700 & & \textbf{0.337} & 0.473 & 0.533 & 0.701 \\
CQA       & \textbf{0.323} & \textbf{0.452} & \textbf{0.452} & \textbf{0.715} & & \textbf{0.338} & \textbf{0.484}& \textbf{0.551} & \textbf{0.710} \\
CQ        & \textbf{0.316} & \textbf{0.448} & \textbf{0.448} & \textbf{0.717} & & 0.334 & \textbf{0.479} & \textbf{0.541} & \textbf{0.705} \\
QA        & 0.307 & 0.437 & 0.437 & 0.706 & & 0.341 & \textbf{0.487} & \textbf{0.550} & \textbf{0.705} \\
A         & 0.304 & 0.424 & 0.424 & 0.696 & & \textbf{0.347} & \textbf{0.493} & \textbf{0.557} & \textbf{0.707} \\
\bottomrule
\end{tabular}
\vspace{2mm}
\caption{Accuracy of Curricular CoT (DEF, CQA, CQ, QA, A) compared with the baseline (Zero-shot)}\label{tab:prompt_accuracy}
\begin{tablenotes}
\footnotesize
\item \textbf{Note:} Bold numbers indicate performance improvements over each model’s zero-shot baseline. Accuracy is averaged across curriculum-competency alignment tasks within each classification granularity level.
\end{tablenotes}
\end{threeparttable}
\end{table}
\subsubsection{Performance heterogeneity across data features}
\begin{figure*}[h]
    \centering
    \begin{subfigure}[t]{0.45\textwidth}
        \centering
        \includegraphics[width=\textwidth]{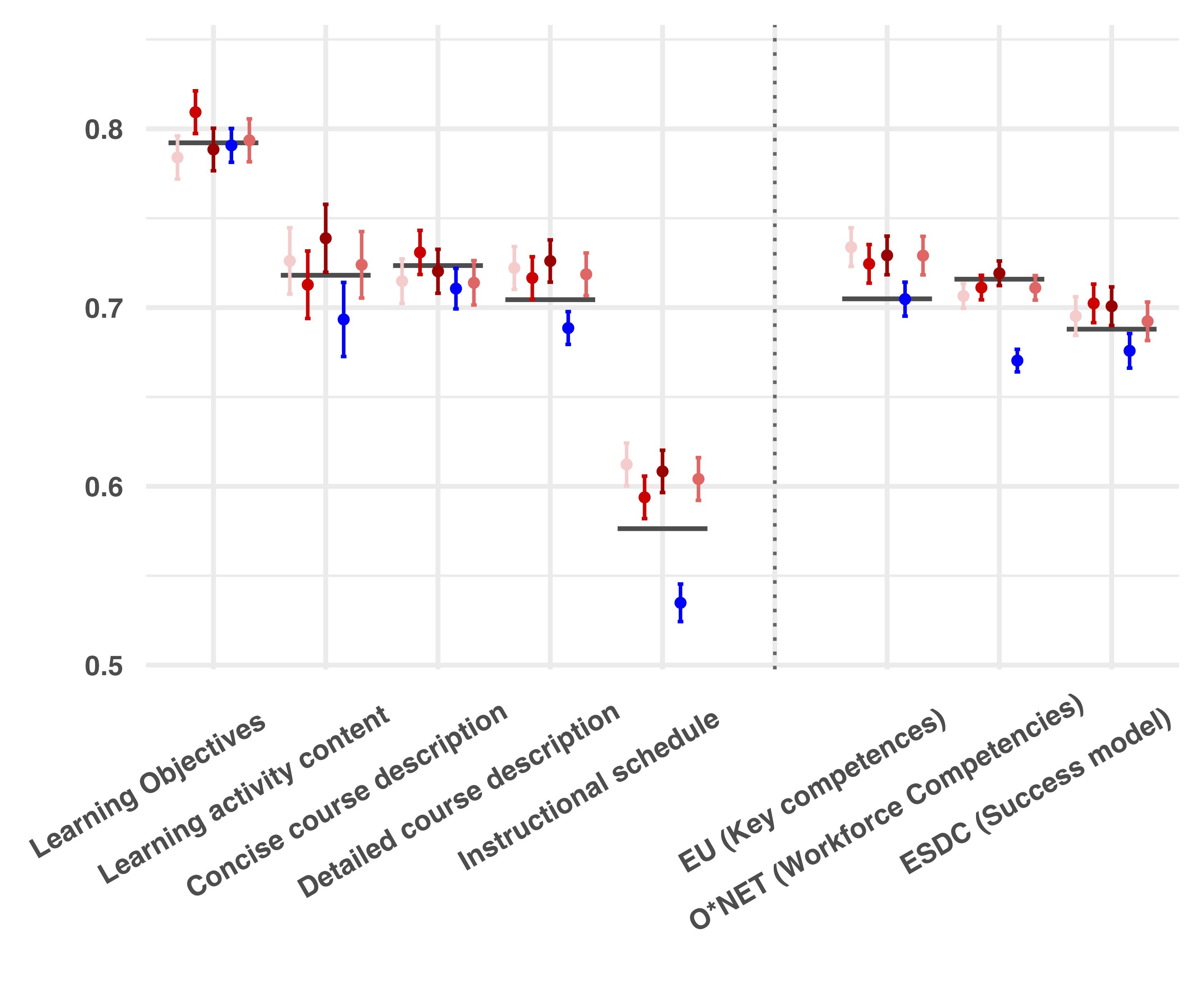}
        \caption{Prediction accuracy (binary classification)}
        \label{fig:RQ3-prediction accuracy}
    \end{subfigure}
    \hfill
    \begin{subfigure}[t]{0.5\textwidth}
        \centering
        \includegraphics[width=\textwidth]{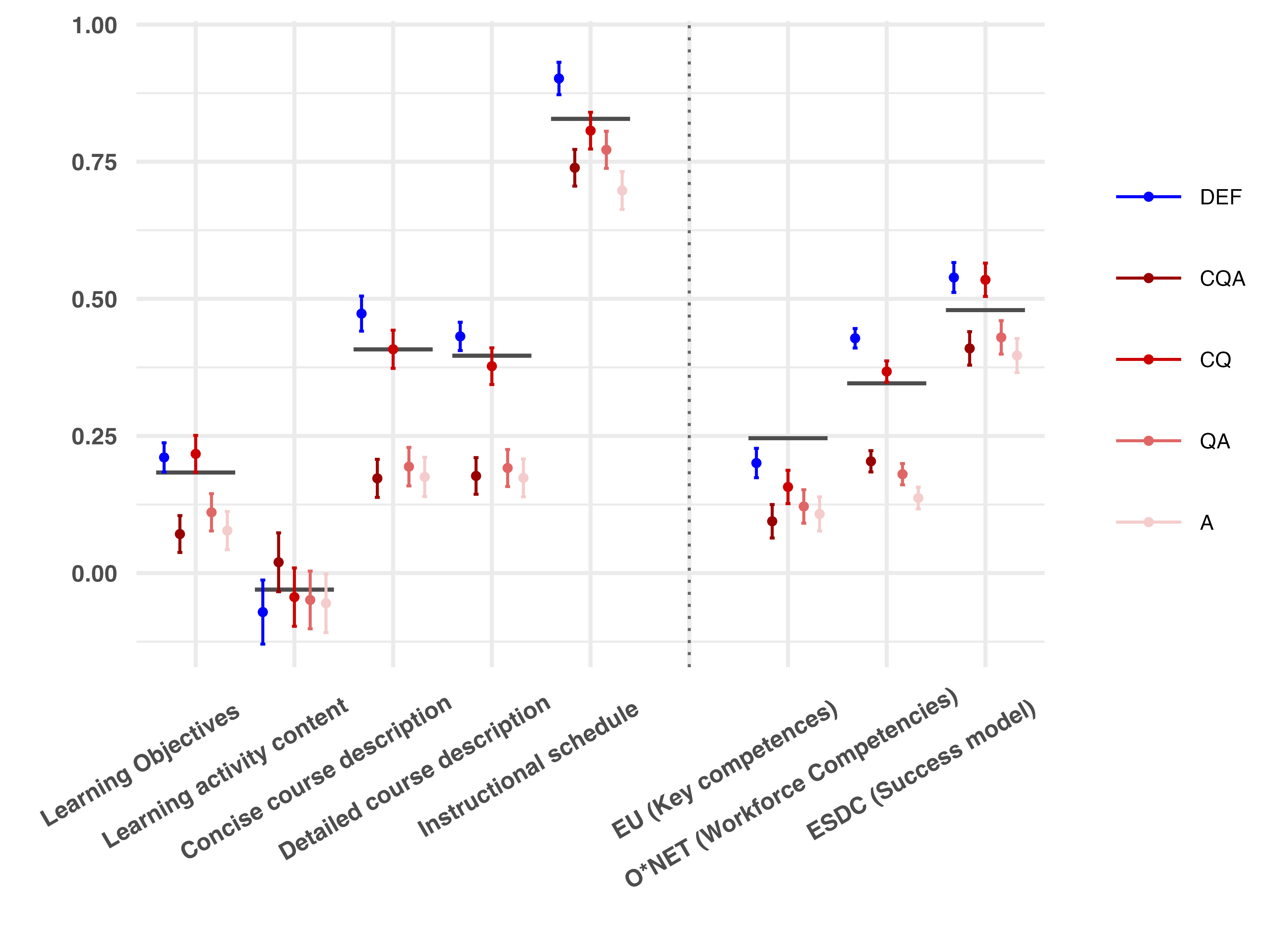}
        \caption{Score differences (5-class classification)} 
        \label{fig:RQ3-score difference}
    \end{subfigure}
    \caption{Estimated performance of Curricular CoT : (a) Accuracy and (b) Score Differences.}
    \begin{tablenotes}
\footnotesize
\item \textbf{Note:} Solid gray lines indicate the estimated accuracy or score difference of the Zero-Shot baseline for each category. Colored points and error bars represent the estimated means and 95\% confidence intervals of other prompting strategies.
\end{tablenotes}
    \label{fig:3}
\end{figure*}

Building on the accuracy gains over the zero-shot baseline, we next examine whether our proposed prompting strategies mitigate the data-specific challenges identified in RQ2. Figure~\ref{fig:RQ3-prediction accuracy} (a) compares the zero-shot baseline with the proposed strategies in binary classification tasks, which we focus on because they are most relevant to practical applications and because prior results show limited zero-shot capability for fine-grained competency analysis.

Overall, Curricular CoT yields modest but consistent improvements in prediction accuracy, particularly by reducing performance gaps across curriculum document types. Notably, for the two document types that performed worst under the zero-shot condition—instructional schedules and detailed course descriptions—each Curricular CoT variant improves accuracy. Even for more information-rich document types, such as learning objectives and learning activity content, several Curricular CoT variants outperform the zero-shot baseline, indicating that structured reasoning can further enhance performance when sufficient instructional detail is available.

In terms of competency frameworks, Curricular CoT methods improve performance for more abstract frameworks, such as the EU (Key Competences) and ESDC (Success Model), but yield limited gains for O*NET (Workforce Competencies). In contrast, the definition-based prompting strategy (DEF) does not reduce performance gaps across frameworks or document types and, in some cases, leads to declines in accuracy.

We next examine a key challenge in the zero-shot setting: LLMs’ tendency to overestimate the coverage of competencies relative to human annotations. Figure~\ref{fig:RQ3-score difference} presents the average differences between LLM-predicted and human rating scores. Because score differences are not meaningful in binary classification, we conduct this analysis in the five-class classification scenario. Our results indicate that CoT-based strategies can mitigate this challenge. These methods consistently reduce the systematic overestimation observed under the zero-shot condition, particularly for inputs with longer word counts and across a range of curriculum document types and competency frameworks. In contrast, the definition-based (DEF) strategy again fails to narrow these discrepancies and, in some cases, further increases the misalignment between LLM-generated scores and human judgments.

\subsubsection{Error analysis} 
\begin{figure*}[htbp]
    \centering
    \includegraphics[width=1\textwidth]{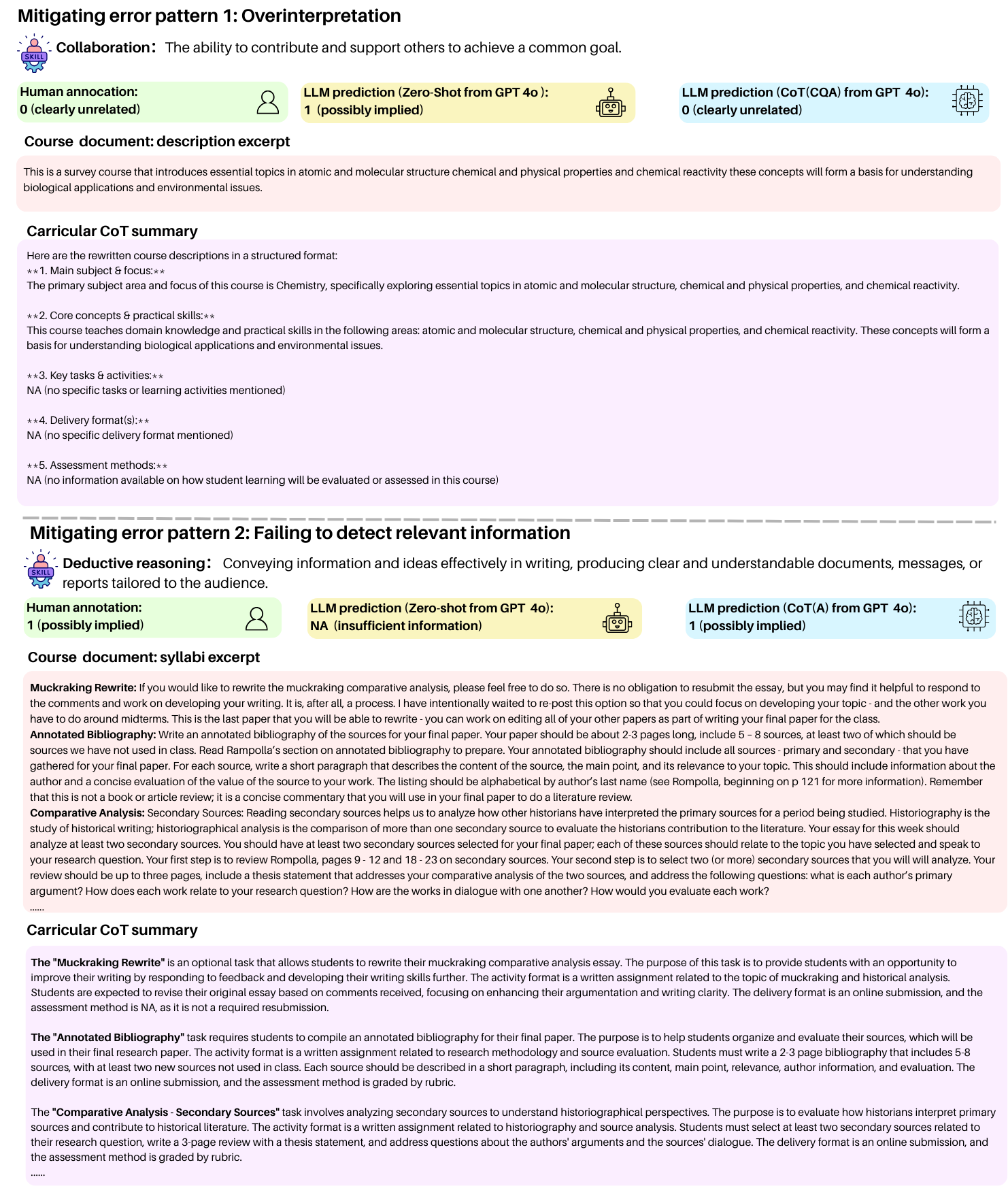}
    \caption{Part 1. Illustrative example: Curricular CoT mitigating off-the-shelf LLM errors}
    \label{fig:error-rq3-1}
\end{figure*}

\begin{figure*}[htbp]
    \centering
    \includegraphics[width=1\textwidth]{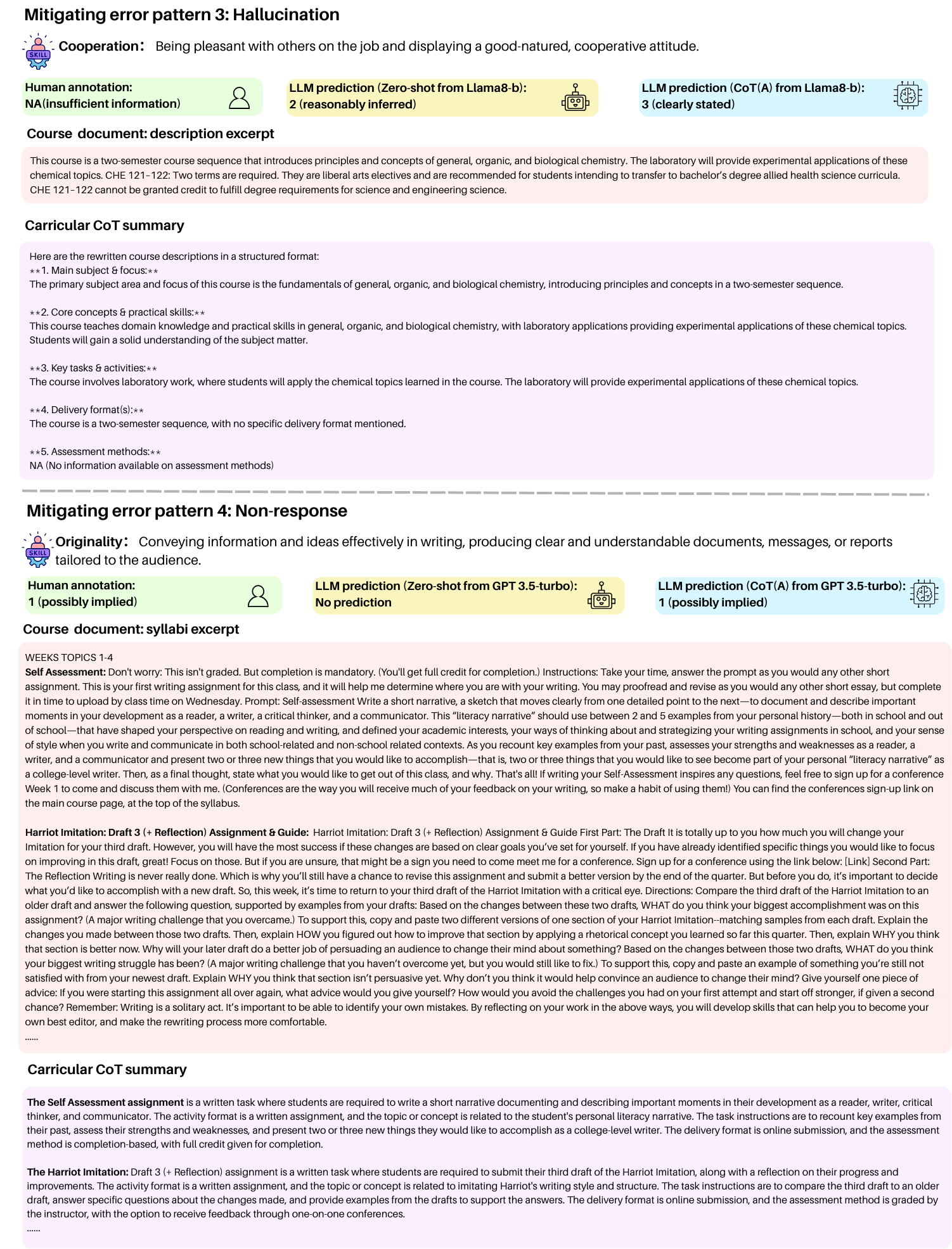}
    \caption{Part 2. Illustrative example: Curricular CoT mitigating off-the-shelf LLM errors}
    \label{fig:error-rq3-2}
\end{figure*}

We conduct a manual error analysis to examine the error patterns identified in the previous section, focusing on how the proposed Curricular CoT approach mitigates these baseline errors and where it still fails. Figures~\ref{fig:error-rq3-1} and~\ref{fig:error-rq3-2} illustrate how the Curricular CoT approach mitigates these errors through intermediate reasoning steps and the structured processing of curriculum documents.

By examining cases in which the Curricular CoT approach mitigates the limitations of direct analysis of curriculum documents, we provide empirical support for our design assumptions. First, consistent with error patterns 1 and 3, when curriculum texts are highly abstract or underspecified, guiding-question-based reasoning enables the model to more accurately extract curriculum elements relevant to competency inference than direct decision making. Particularly when relevant information is missing, the model provides a more honest summary of what is available in the curriculum document and explicitly reports the lack of information instead of making overconfident assumptions, as seen in the direct inference of higher-order skills. Moreover, in error patterns 2 and 4 involving lengthy curriculum documents, the Curricular CoT approach facilitates the identification of subtle yet meaningful instructional cues. By structuring the extraction process around targeted guiding questions, the model is better able to surface relevant information embedded in dense descriptions, thereby reducing the likelihood of missing evidence in long texts.

We also analyze failure cases in which the proposed methods yield limited or negative gains. A primary source of failure is variability in the quality of intermediate summaries generated by different LLMs. Because Curricular CoT relies on the accurate extraction and structuring of curriculum content prior to competency inference, errors in summarization can degrade downstream performance. Comparing intermediate summaries across models reveals clear differences. GPT-4o produces the most accurate and well-structured summaries, consistently capturing key instructional components. Llama3-8B generates concise but less detailed summaries. In contrast, GPT-3.5-turbo frequently hallucinates unsupported instructional elements, while Llama3-70B, although generally accurate, often introduces redundancy that adds noise to downstream analysis.

\section{Discussion}
In this study, we investigate the affordances of large language models for analyzing the integration of 21st-century competencies in postsecondary education curricula using digital curriculum documents commonly available at postsecondary institutions. We evaluate the usability of these data sources, benchmark zero-shot LLM performance, and identify systematic reasoning errors. To address the lack of large-scale, high-quality curriculum benchmarks, we propose a reasoning-based prompting strategy that improves LLMs’ pedagogically grounded reasoning over complex and unstructured curriculum texts.

Regarding data feasibility, our findings suggest that digital curriculum documents in postsecondary institutions constitute a valuable yet underutilized resource for examining how 21st-century competencies are embedded within courses and programs. A key concern in prior literature is whether such data contain sufficient pedagogical detail to support meaningful analysis. From a theoretical perspective, the development of 21st-century competencies typically relies on student-centered instructional designs, such as project-based and active learning ~\citep{meyers2009use, musa2012project}, which require curriculum documentation with adequate informational richness. Our human examination confirms that learning activity descriptions are the most informative and reliable data source for competency assessment and should be prioritized when available. Learning management systems offer a promising source for capturing such detailed records. In contrast, instructional schedules in syllabi, while also indicating activity types, often lack content-level specificity and are less feasible for analyzing 21st-century competencies. Course descriptions and learning goals, typically drawn from syllabi or institutional websites, showed moderate utility but varied widely in quality across institutions. Importantly, these findings are not intended to prescribe a fixed hierarchy among curriculum document types. Rather, we argue that the presence of explicit pedagogical elements is more critical than the source of the curriculum document in determining suitability for competency analytics. In practice, institutions may benefit from combining multiple data types to achieve sufficient pedagogical coverage, tailored to the characteristics and informational richness of their local data ecosystems. Over the long term, richer and more structured curriculum data infrastructures will be essential for enabling systematic curricular analytics to support monitoring, evaluation, and continuous improvement.

In terms of the affordances of current general-purpose LLMs for higher-order competency analytics, our results reveal both promise and clear limitations. Even the most advanced model in our study (GPT-4o) does not yet achieve the analytical precision of human experts. While human annotators can reliably distinguish subtle differences in the extent to which courses integrate specific competencies, LLMs struggle with such fine-grained distinctions, highlighting the challenge of applying general-purpose models to pedagogical judgments that require contextual interpretation. Nonetheless, LLMs perform reasonably well in coarse-grained tasks, such as identifying whether a course likely addresses a given competency. The relatively small performance gap between large proprietary models and open-weight models further suggests the cost-effectiveness and scalability of open-weight models for large-scale curricular analytics. We further identify two main sources of model error. First, when analyzing abstract course data, LLMs tend to over-infer or hallucinate based on some keywords (e.g., assuming any “final exam” implies writing comprehension), likely due to patterns learned from historical training data. Second, when processing long and unstructured course text, models often fail to locate the precise pedagogical evidence needed for accurate inference. This issue may reflect constraints in contextual processing and attention span. Together, these findings provide empirical insight into why LLMs continue to struggle with complex educational reasoning tasks, consistent with recent work ~\citep{xu2025course,siyan2025bringing}. They also motivate our element-centric prompting strategy, which aims to strengthen in-context pedagogical reasoning by explicitly directing models’ attention to instructional elements most relevant to competency integration.

As prompt engineering remains the most practical approach for LLM-supported curricular analytics, we evaluate our proposed prompting strategies across multiple curriculum document types, competency frameworks, and both open-weight and proprietary models. Our results show that the CoT-based approach yields modest performance gains, particularly for more advanced models. Qualitative analysis suggests that these gains arise because the guided extraction of structured instructional elements enables more accurate inference of competencies than direct reasoning from raw curriculum text. This intermediate step mitigates hallucination in abstract texts and improves the identification of relevant pedagogical evidence in lengthy, unstructured content. Consistent with prior CoT research, decomposing complex reasoning into explicit steps enhances performance, though the effectiveness of this approach depends on the quality of the intermediate extraction. When reasoning quality is low, this step can introduce noise and reduce accuracy, making the method most effective for larger, more capable models ~\citep{wei2022emergent}. Interestingly, incorporating human-defined competency descriptions into prompts does not improve LLM performance. This finding supports our hypothesis that higher-order competency analytics require more than surface-level definition matching and instead depend on pedagogically grounded reasoning. Human experts, when performing this task, frequently draw on implicit knowledge when evaluating whether an activity meaningfully promotes a given skill, rather than relying solely on predefined criteria. Such pedagogical knowledge is difficult to formalize or encode in prompts for general-purpose LLMs, limiting their effectiveness. These results underscore the constraints of current LLMs and highlight the need for models or training strategies specifically designed to support pedagogical reasoning in curricular analytics.

Finally, there are still several limitations in our current study that could be addressed in future work. First, our analysis is conducted on a limited set of annotated curriculum samples. Although we conduct human checks to ensure their representativeness, future research should draw on more diverse and multiple data sources to improve generalization. Second, the LLM prompts in this study were not iteratively refined using a separate training or validation set. Future work could incorporate such a step to further optimize prompt design in the specific institutional context. Third, while our evaluation approach allows us to isolate and better understand the limitations and strengths of each data type, it does not reflect how institutions typically use curriculum documents in practice. In real-world settings, multiple document types are often used together. Future research should therefore examine how integrating multiple data sources affects LLM performance and whether such integration can improve robustness and reliability.

Beyond addressing these limitations, our findings point to several directions for future research in LLM-assisted curricular analytics. The lack of large-scale, high-quality benchmark datasets remains a major bottleneck for curricular analytics. Unlike domains such as the labor market, where richly annotated corpora have accelerated methodological progress, postsecondary curricular analytics lacks shared benchmarks. Future work should prioritize the development and open sharing of such datasets, potentially through coordinated efforts in structured data collection and automated or semi-automated annotation of existing curriculum documents ~\citep{javadian2024course}. Moreover, our results indicate that the effectiveness of reasoning-based prompting depends strongly on model capacity. Hybrid pipelines that combine models of different sizes—for example, using more powerful models for content extraction and lighter models for scoring—may offer a practical trade-off between performance and computational cost. 

\section{Conclusion}
In this study, we examine whether and how digital curriculum documents, commonly available at postsecondary institutions, can support the analysis of 21st-century competency integration and evaluate the extent to which LLMs can reliably perform this task. Using three competency frameworks and five types of curriculum documents, we manually annotated 7,600 curriculum-competency alignments to assess data feasibility, benchmark zero-shot LLM performance, and test a reasoning-based prompting strategy.

We first show that existing curriculum documents constitute a valuable but underutilized resource for competency-oriented analytics. Among the document types examined, detailed descriptions of learning activities provide the most informative signals for assessing competency integration. More importantly, effective analysis depends on whether curriculum documents contain explicit pedagogical elements, suggesting that combining multiple document types is often necessary to achieve adequate informational coverage in institutional settings.

We then demonstrate that current general-purpose LLMs have mixed capabilities for higher-order competency analytics. While both proprietary and open-weight models perform well in coarse-grained classifications, achieving over 70\% agreement with human annotations, they fall short of human-level performance in fine-grained assessments of competency coverage. We identify two primary sources of error: systematic over-interpretation of instructional keywords and difficulty identifying relevant pedagogical evidence in long, unstructured texts.

To address these limitations, we introduce a reasoning-based prompting strategy that guides models to extract and reason over structured instructional elements before making judgments. This approach yields modest but consistent performance improvements across data types and competency frameworks, consistent with the Chain-of-Thought principle that decomposing complex reasoning tasks enhances model performance. However, its effectiveness depends on model capacity, as errors in intermediate extraction can introduce noise and reduce analytical accuracy.

Taken together, our work extends prior evaluations of LLMs for curricular analytics by systematically examining data feasibility, model performance, and technical interventions across multiple data types and frameworks. Our findings indicate that realizing the full potential of curricular analytics requires not only technical advances but also higher-quality curriculum data infrastructures and shared benchmark datasets. While general-purpose LLMs remain limited in their ability to perform complex pedagogical reasoning, their performance can be improved through structured reasoning designs and careful data use. Future gains are likely to come from richer and more standardized curriculum data, shared benchmark datasets, and hybrid human-AI or domain-specific modeling approaches that better align with the pedagogical nature of curricular analytics.

\phantomsection
\section*{Acknowledgments} 
We thank Joe Karaganis for providing access to Open Syllabus data.
\addcontentsline{toc}{section}{Acknowledgments} 

\section*{Declaration of Conflicting Interest} 
\addcontentsline{toc}{section}{Declaration of Conflicting Interest} 
The authors declared no potential conflicts of interest with respect to the research, authorship, and/or publication of this article.
\section*{Funding} 
The author(s) declared no financial support for the research, authorship, and/or publication of this article

\addcontentsline{toc}{section}{Funding} 

\phantomsection
\printbibliography

@article{de2024curriculum,
	title = {Curriculum analytics in higher education institutions: a systematic literature review},
	author = {De Silva, Liyanachchi Mahesha Harshani and Rodr{\'\i}guez-Triana, Mar{\'\i}a Jes{\'u}s and Chounta, Irene-Angelica and Pishtari, Gerti},
	journal = {Journal of Computing in Higher Education},
	pages = {1--47},
	publisher = {Springer},
	year = {2024},
}

@article{wei2022chain,
	title = {Chain-of-thought prompting elicits reasoning in large language models},
	author = {Wei, Jason and Wang, Xuezhi and Schuurmans, Dale and Bosma, Maarten and Xia, Fei and Chi, Ed and Le, Quoc V and Zhou, Denny and others},
	journal = {Advances in neural information processing systems},
	pages = {24824--24837},
	volume = {35},
	year = {2022},
}

@article{wang2023element,
	title = {Element-aware summarization with large language models: Expert-aligned evaluation and chain-of-thought method},
	author = {Wang, Yiming and Zhang, Zhuosheng and Wang, Rui},
	journal = {arXiv preprint arXiv:2305.13412},
	year = {2023},
}

@article{xu2025course,
	title = {From Course to Skill: Evaluating LLM Performance in Curricular Analytics},
	author = {Xu, Zhen and Li, Xinjin and Huan, Yingqi and Minaya, Veronica and Yu, Renzhe},
	journal = {arXiv preprint arXiv:2505.02324},
	year = {2025},
}

@inproceedings{zamecnik2024mapping,
	title = {Mapping Employable Skills in Higher Education Curriculum Using LLMs},
	author = {Zamecnik, Andrew and Barthakur, Abhinava and Wang, Hanyi and Dawson, Shane},
	booktitle = {European Conference on Technology Enhanced Learning},
	organization = {Springer},
	pages = {18--32},
	year = {2024},
}

@article{dawson2014curriculum,
	title = {Curriculum analytics: Application of social network analysis for improving strategic curriculum decision-making in a research-intensive university},
	author = {Dawson, Shane and Hubball, Harry},
	journal = {Teaching and Learning Inquiry},
	number = {2},
	pages = {59--74},
	publisher = {JSTOR},
	volume = {2},
	year = {2014},
}

@article{jovanovic2025curriculum,
	title = {Curriculum analytics: Exploring assessment objectives, types, and grades in a study program},
	author = {Jovanovi{\'c}, Jelena and Zamecnik, Andrew and Barthakur, Abhinava and Dawson, Shane},
	journal = {Education and Information Technologies},
	number = {4},
	pages = {4843--4866},
	publisher = {Springer},
	volume = {30},
	year = {2025},
}

@inproceedings{jayalath2025scaling,
	title = {Scaling Curriculum Mapping in Higher Education: Evaluating Generative AI’s Role in Curriculum Analytics},
	author = {Jayalath, Vimukthini and Barthakur, Abhinava and Dawson, Shane and Tingey, Joanne and Crase, Lin and Kovanovi{\'c}, Vitomir},
	booktitle = {International Conference on Artificial Intelligence in Education},
	organization = {Springer},
	pages = {294--308},
	year = {2025},
}

@article{sridhar2023harnessing,
	title = {Harnessing llms in curricular design: Using gpt-4 to support authoring of learning objectives},
	author = {Sridhar, Pragnya and Doyle, Aidan and Agarwal, Arav and Bogart, Christopher and Savelka, Jaromir and Sakr, Majd},
	journal = {arXiv preprint arXiv:2306.17459},
	year = {2023},
}

@article{siyan2025bringing,
	title = {Bringing Pedagogy into Focus: Evaluating Virtual Teaching Assistants' Question-Answering in Asynchronous Learning Environments},
	author = {Siyan, Li and Xu, Zhen and Raghuram, Vethavikashini Chithrra and Zhang, Xuanming and Yu, Renzhe and Yu, Zhou},
	journal = {arXiv preprint arXiv:2509.17961},
	year = {2025},
}

@article{retnawati2018teachers,
	title = {Teachers' Knowledge about Higher-Order Thinking Skills and Its Learning Strategy.},
	author = {Retnawati, Heri and Djidu, Hasan and Apino, Ezi and Anazifa, Risqa D and others},
	journal = {Problems of Education in the 21st Century},
	number = {2},
	pages = {215--230},
	publisher = {Scientia Socialis Ltd. 29 K. Donelaicio Street, LT-78115 Siauliai, Republic~…},
	volume = {76},
	year = {2018},
}

@inproceedings{shorman2024curriculum,
	title = {Curriculum Management System to Measure the Course and Program Outcomes},
	author = {Shorman, Samer and Khder, Moaiad and others},
	booktitle = {2024 ASU International Conference in Emerging Technologies for Sustainability and Intelligent Systems (ICETSIS)},
	organization = {IEEE},
	pages = {391--397},
	year = {2024},
}

@incollection{ghanizadeh2020higher,
	title = {Higher order thinking skills},
	author = {Ghanizadeh, Afsaneh and Al-Hoorie, Ali H and Jahedizadeh, Safoura},
	booktitle = {Higher Order Thinking Skills in the Language Classroom: A Concise Guide},
	pages = {1--51},
	publisher = {Springer},
	year = {2020},
}

@book{griffin2012assessment,
	title = {Assessment and teaching of 21st century skills},
	author = {Griffin, Patrick and McGaw, Barry and Care, Esther},
	publisher = {Springer},
	volume = {10},
	year = {2012},
}

@article{kotsiou2022scoping,
	title = {A scoping review of Future Skills frameworks},
	author = {Kotsiou, Athanasia and Fajardo-Tovar, Dina Daniela and Cowhitt, Tom and Major, Louis and Wegerif, Rupert},
	journal = {Irish Educational Studies},
	number = {1},
	pages = {171--186},
	publisher = {Taylor \& Francis},
	volume = {41},
	year = {2022},
}

@article{nguyen2024rethinking,
	title = {Rethinking skill extraction in the job market domain using large language models},
	author = {Nguyen, Khanh Cao and Zhang, Mike and Montariol, Syrielle and Bosselut, Antoine},
	journal = {arXiv preprint arXiv:2402.03832},
	year = {2024},
}

@article{senger2024deep,
	title = {Deep learning-based computational job market analysis: A survey on skill extraction and classification from job postings},
	author = {Senger, Elena and Zhang, Mike and van der Goot, Rob and Plank, Barbara},
	journal = {arXiv preprint arXiv:2402.05617},
	year = {2024},
}

@article{herandi2024skill,
	title = {Skill-LLM: Repurposing General-Purpose LLMs for Skill Extraction},
	author = {Herandi, Amirhossein and Li, Yitao and Liu, Zhanlin and Hu, Ximin and Cai, Xiao},
	journal = {arXiv preprint arXiv:2410.12052},
	year = {2024},
}

@article{myronenko2024improving,
	title = {Improving Skill Extraction from Job Postings Using Synthetic Data and Advanced Language Models},
	author = {Myronenko, Andrii},
	year = {2024},
}

@article{thakrar2025enhancing,
	title = {Enhancing Talent Employment Insights Through Feature Extraction with LLM Finetuning},
	author = {Thakrar, Karishma and Young, Nick},
	journal = {arXiv preprint arXiv:2501.07663},
	year = {2025},
}

@article{buckingham2016learning,
	title = {Learning Analytics for 21st Century Competencies.},
	author = {Buckingham Shum, Simon and Crick, Ruth Deakin},
	journal = {Journal of Learning Analytics},
	number = {2},
	pages = {6--21},
	publisher = {ERIC},
	volume = {3},
	year = {2016},
}

@article{hilliger2024curriculum,
	title = {Curriculum analytics adoption in higher education: A multiple case study engaging stakeholders in different phases of design},
	author = {Hilliger, Isabel and Miranda, Constanza and Celis, Sergio and P{\'e}rez-Sanagust{\'\i}n, Mar},
	journal = {British Journal of Educational Technology},
	number = {3},
	pages = {785--801},
	publisher = {Wiley Online Library},
	volume = {55},
	year = {2024},
}

@inproceedings{greer2016learning,
	title = {Learning analytics for curriculum and program quality improvement (PCLA 2016)},
	author = {Greer, Jim and Molinaro, Marco and Ochoa, Xavier and McKay, Timothy},
	booktitle = {Proceedings of the sixth international conference on learning analytics \& knowledge},
	pages = {494--495},
	year = {2016},
}

@inproceedings{ohland2002creating,
	title = {Creating A Catalog And Meta Analysis Of Freshman Programs For Engineering Students: Part 2: Learning Communities},
	author = {Ohland, Matthew and Collins, Rachel},
	booktitle = {2002 Annual Conference},
	pages = {7--338},
	year = {2002},
}

@inproceedings{russell2024mapping,
	title = {Mapping Curricula to Skills and Occupations Using Course Descriptions},
	author = {Russell E, Walker},
	booktitle = {2024 IEEE World Engineering Education Conference (EDUNINE)},
	organization = {IEEE},
	pages = {1--6},
	year = {2024},
}

@article{yang2023analyzing,
	title = {Analyzing the alignment between AI curriculum and AI textbooks through text mining},
	author = {Yang, Hyeji and Kim, Jamee and Lee, Wongyu},
	journal = {Applied Sciences},
	number = {18},
	pages = {10011},
	publisher = {MDPI},
	volume = {13},
	year = {2023},
}

@article{javadian2024course,
	title = {Course-Skill Atlas: A national longitudinal dataset of skills taught in US higher education curricula},
	author = {Javadian Sabet, Alireza and Bana, Sarah H and Yu, Renzhe and Frank, Morgan R},
	journal = {Scientific Data},
	number = {1},
	pages = {1086},
	publisher = {Nature Publishing Group UK London},
	volume = {11},
	year = {2024},
}

@article{light2024student,
	title = {Student demand and the supply of college courses},
	author = {Light, Jacob},
	journal = {Available at SSRN 4856488},
	year = {2024},
}

@article{nye2021show,
	title = {Show Your Work: Scratchpads for Intermediate Computation with Language Models},
	author = {Nye, Max and Hewitt, John and Chen, Jiajun and Krueger, David and Duvenaud, David and Lake, Brenden and Zemel, Richard},
	journal = {arXiv preprint arXiv:2112.00114},
	year = {2021},
}

@article{wei2022emergent,
	title = {Emergent abilities of large language models},
	author = {Wei, Jason and Tay, Yi and Bommasani, Rishi and Raffel, Colin and Zoph, Barret and Borgeaud, Sebastian and Yogatama, Dani and Bosma, Maarten and Zhou, Denny and Metzler, Donald and others},
	journal = {arXiv preprint arXiv:2206.07682},
	year = {2022},
}

@article{kojima2022large,
	title = {Large language models are zero-shot reasoners},
	author = {Kojima, Takeshi and Gu, Shixiang Shane and Reid, Machel and Matsuo, Yutaka and Iwasawa, Yusuke},
	journal = {arXiv preprint arXiv:2205.11916},
	year = {2022},
}

@inproceedings{zhang2022automatic,
	title = {Automatic chain of thought prompting in large language models},
	author = {Zhang, Shinn and Qin, Lianmin and Zhou, Denny and Le, Quoc V and Liu, Peter J and others},
	booktitle = {arXiv preprint arXiv:2210.03493},
	year = {2022},
}

@article{shao2023synthetic,
	title = {Synthetic prompting: Chain-of-thought prompting with generated questions and answers},
	author = {Shao, Yujia and Shao, Zhouxing and Xu, Canwen and others},
	journal = {arXiv preprint arXiv:2302.01348},
	year = {2023},
}

@inproceedings{zhou2022least,
	title = {Least-to-most prompting enables complex reasoning in large language models},
	author = {Zhou, Denny and Sch{\"a}rli, Nathanael and Hou, Le and Wei, Jason and Wang, Xuezhi and Singh, Amanpreet and Thoppilan, Romal and Chung, Hyung Won and Chung, Mia and Menegaux, Paul and others},
	booktitle = {International Conference on Learning Representations (ICLR)},
	year = {2022},
}

@article{yao2023tree,
	title = {Tree of thoughts: Deliberate problem solving with large language models},
	author = {Yao, Shinn and Zhao, Jinhua and Yu, Dian and others},
	journal = {arXiv preprint arXiv:2305.10601},
	year = {2023},
}

@article{deng2024rephrase,
	title = {Rephrase and Respond: Let Large Language Models Ask Better Questions for Themselves},
	author = {Deng, Yihe and Zhang, Weitong and Chen, Zixiang and Gu, Quanquan},
	journal = {arXiv preprint arXiv:2311.04205},
	year = {2024},
}

@book{oecd2023employment,
	title = {OECD Employment Outlook 2023: Artificial Intelligence and the Labour Market},
	author = {{OECD}},
	address = {Paris},
	doi = {10.1787/08785bba-en},
	publisher = {OECD Publishing},
	url = {https://doi.org/10.1787/08785bba-en},
	year = {2023},
}

@book{oecd2023,
	title = {Innovating Assessments to Measure and Support Complex Skills},
	author = {{OECD}},
	address = {Paris},
	doi = {10.1787/e5f3e341-en},
	editor = {Naomi Foster and Mario Piacentini},
	note = {Edited by Naomi Foster and Mario Piacentini},
	publisher = {OECD Publishing},
	url = {https://doi.org/10.1787/e5f3e341-en},
	year = {2023},
}

@report{wef2025,
	title = {The Future of Jobs Report 2025},
	author = {{World Economic Forum}},
	address = {Geneva},
	institution = {World Economic Forum},
	note = {Accessed: 2025-06-07},
	url = {https://www.weforum.org/publications/the-future-of-jobs-report-2025/},
	year = {2025},
}

@techreport{used2023ai,
	title = {Artificial Intelligence and the Future of Teaching and Learning: Insights and Recommendations},
	author = {{U.S. Department of Education}},
	address = {Washington, DC},
	note = {Accessed: 2025-06-07},
	url = {https://www.ed.gov/AI},
	year = {2023},
}

@techreport{unesco2023genai,
	title = {Guidance for Generative AI in Education and Research},
	author = {UNESCO},
	address = {Paris},
	note = {Accessed: 2025-06-07},
	url = {https://unesdoc.unesco.org/ark:/48223/pf0000386693},
	year = {2023},
}

@misc{ukgenerative,
	title = {Generative Artificial Intelligence (AI) in Education},
	author = {{UK GOV}},
	howpublished = {\url{https://www.gov.uk/government/publications/generative-artificial-intelligence-in-education}},
	note = {Accessed: 2025-06-07},
	year = {2023},
}

@techreport{nea2024ai,
	title = {Teaching in the Age of AI: NEA Members' Roadmap for Safe, Effective, and Accessible Use of Artificial Intelligence in Education},
	author = {{National Education Association}},
	address = {Washington, DC},
	note = {Accessed: 2025-06-07},
	url = {https://www.nea.org/resource-library/artificial-intelligence-education},
	year = {2024},
}

@book{oecd2023digital,
	title = {OECD Digital Education Outlook 2023: Towards an Effective Digital Education Ecosystem},
	author = {{OECD}},
	address = {Paris},
	doi = {10.1787/c74f03de-en},
	note = {Accessed: 2025-06-07},
	publisher = {OECD Publishing},
	url = {https://doi.org/10.1787/c74f03de-en},
	year = {2023},
}

@report{ellingrud2023generative,
	title = {Generative AI and the Future of Work in America},
	author = {{McKinsey Global Institute}},
	month = {July},
	url = {https://www.mckinsey.com/mgi/our-research/generative-ai-and-the-future-of-work-in-america},
	year = {2023},
}

@article{pistilli2017guiding,
	title = {Guiding early and often: Using curricular and learning analytics to shape teaching, learning, and student success in gateway courses},
	author = {Pistilli, Matthew D and Heileman, Gregory L},
	journal = {New Directions for Higher Education},
	number = {180},
	pages = {21--30},
	publisher = {Wiley Online Library},
	volume = {2017},
	year = {2017},
}

@article{chou2015open,
	title = {Open student models of core competencies at the curriculum level: Using learning analytics for student reflection},
	author = {Chou, Chih-Yueh and Tseng, Shu-Fen and Chih, Wen-Chieh and Chen, Zhi-Hong and Chao, Po-Yao and Lai, K Robert and Chan, Chien-Lung and Yu, Liang-Chih and Lin, Yi-Lung},
	journal = {IEEE Transactions on Emerging Topics in Computing},
	number = {1},
	pages = {32--44},
	publisher = {IEEE},
	volume = {5},
	year = {2015},
}

@inproceedings{hilliger2020design,
	title = {Design of a curriculum analytics tool to support continuous improvement processes in higher education},
	author = {Hilliger, Isabel and Aguirre, Camila and Miranda, Constanza and Celis, Sergio and P{\'e}rez-Sanagust{\'\i}n, Mar},
	booktitle = {Proceedings of the tenth international conference on learning analytics \& knowledge},
	pages = {181--186},
	year = {2020},
}

@article{irwin2002characterizing,
	title = {Characterizing the core: What catalog descriptions of mandatory courses reveal about LIS schools and librarianship},
	author = {Irwin, Ray},
	journal = {Journal of Education for Library and Information Science},
	pages = {175--184},
	publisher = {JSTOR},
	year = {2002},
}

@inproceedings{fiesler2020we,
	title = {What do we teach when we teach tech ethics? A syllabi analysis},
	author = {Fiesler, Casey and Garrett, Natalie and Beard, Nathan},
	booktitle = {Proceedings of the 51st ACM technical symposium on computer science education},
	pages = {289--295},
	year = {2020},
}

@article{gorski2009we,
	title = {What we're teaching teachers: An analysis of multicultural teacher education coursework syllabi},
	author = {Gorski, Paul C},
	journal = {Teaching and teacher education},
	number = {2},
	pages = {309--318},
	publisher = {Elsevier},
	volume = {25},
	year = {2009},
}

@inproceedings{li2024,
	title = {Supporting Teaching-to-the-Curriculum by Linking Diagnostic Tests to Curriculum Goals},
	author = {Li, X. and Henriksson, A. and Duneld, M. and Nouri, J. and Wu, Y.},
	booktitle = {Artificial Intelligence in Education},
	doi = {10.1007/978-3-031-64302-6_9},
	pages = {118--132},
	publisher = {Springer Nature Switzerland},
	volume = {14829},
	year = {2024},
}

@article{homa2013analysis,
	title = {An analysis of learning objectives and content coverage in introductory psychology syllabi},
	author = {Homa, Natalie and Hackathorn, Jana and Brown, Carrie M and Garczynski, Amy and Solomon, Erin D and Tennial, Rachel and Sanborn, Ursula A and Gurung, Regan AR},
	journal = {Teaching of Psychology},
	number = {3},
	pages = {169--174},
	publisher = {Sage Publications Sage CA: Los Angeles, CA},
	volume = {40},
	year = {2013},
}

@article{hong2009understanding,
	title = {Understanding social justice in social work: A content analysis of course syllabi},
	author = {Hong, Philip Young P and Hodge, David R},
	journal = {Families in Society},
	number = {2},
	pages = {212--219},
	publisher = {SAGE Publications Sage CA: Los Angeles, CA},
	volume = {90},
	year = {2009},
}

@article{tang2016data,
	title = {Data science programs in US higher education: An exploratory content analysis of program description, curriculum structure, and course focus},
	author = {Tang, Rong and Sae-Lim, Watinee},
	journal = {Education for Information},
	number = {3},
	pages = {269--290},
	publisher = {IOS Press},
	volume = {32},
	year = {2016},
}

@article{zhang2022skillspan,
	title = {SkillSpan: Hard and soft skill extraction from English job postings},
	author = {Zhang, Mike and Jensen, Kristian N{\o}rgaard and Sonniks, Sif Dam and Plank, Barbara},
	journal = {arXiv preprint arXiv:2204.12811},
	year = {2022},
}

@article{doyle2025,
	title = {A comparative study of {AI}‐generated and human‐crafted learning objectives in computing education},
	author = {Doyle, A. and Sridhar, P. and Agarwal, A. and Savelka, J. and Sakr, M.},
	doi = {10.1111/jcal.13092},
	journal = {Journal of Computer Assisted Learning},
	number = {1},
	pages = {e13092},
	volume = {41},
	year = {2025},
}

@inproceedings{kawintiranon2016,
	title = {Understanding knowledge areas in curriculum through text mining from course materials},
	author = {Kawintiranon, K. and Vateekul, P. and Suchato, A. and Punyabukkana, P.},
	booktitle = {2016 IEEE International Conference on Teaching, Assessment, and Learning for Engineering (TALE)},
	doi = {10.1109/TALE.2016.7851788},
	pages = {161--168},
	year = {2016},
}

@article{kozov2024,
	title = {Practical Application of {AI} and Large Language Models in Software Engineering Education},
	author = {Kozov, V. and Ivanova, G. and Atanasova, D.},
	doi = {10.14569/IJACSA.2024.0150168},
	journal = {International Journal of Advanced Computer Science and Applications},
	number = {1},
	volume = {15},
	year = {2024},
}

@inproceedings{tan2023,
	title = {Revolutionizing Formative Assessment in {STEM} Fields: Leveraging {AI} and {NLP} Techniques},
	author = {Tan, C. W. and Lim, K. Y.},
	booktitle = {2023 Asia Pacific Signal and Information Processing Association Annual Summit and Conference (APSIPA ASC)},
	doi = {10.1109/APSIPAASC58517.2023.10317226},
	pages = {1357--1364},
	year = {2023},
}

@inproceedings{liu2024hita,
	title = {{HiTA}: A {RAG}-Based Educational Platform that Centers Educators in the Instructional Loop},
	author = {Liu, Chang and Hoang, Loc and Stolman, Andrew and Wu, Bo},
	booktitle = {International Conference on Artificial Intelligence in Education},
	organization = {Springer},
	pages = {405--412},
	year = {2024},
}

@article{lohr2025leveraging,
	title = {Leveraging Large Language Models to Generate Course-Specific Semantically Annotated Learning Objects},
	author = {Lohr, Dominic and Berges, Marc and Chugh, Abhishek and Kohlhase, Michael and M{\"u}ller, Dennis},
	journal = {Journal of Computer Assisted Learning},
	number = {1},
	pages = {e13101},
	publisher = {Wiley Online Library},
	volume = {41},
	year = {2025},
}

@inproceedings{lyu2024evaluating,
	title = {Evaluating the effectiveness of llms in introductory computer science education: A semester-long field study},
	author = {Lyu, Wenhan and Wang, Yimeng and Chung, Tingting and Sun, Yifan and Zhang, Yixuan},
	booktitle = {Proceedings of the Eleventh ACM Conference on Learning@ Scale},
	pages = {63--74},
	year = {2024},
}

@inproceedings{kitto2020towards,
	title = {Towards skills-based curriculum analytics: Can we automate the recognition of prior learning?},
	author = {Kitto, Kirsty and Sarathy, Nikhil and Gromov, Aleksandr and Liu, Ming and Musial, Katarzyna and Buckingham Shum, Simon},
	booktitle = {Proceedings of the tenth international conference on learning analytics \& knowledge},
	pages = {171--180},
	year = {2020},
}

@article{decorte2022design,
	title = {Design of negative sampling strategies for distantly supervised skill extraction},
	author = {Decorte, Jens-Joris and Van Hautte, Jeroen and Deleu, Johannes and Develder, Chris and Demeester, Thomas},
	journal = {arXiv preprint arXiv:2209.05987},
	year = {2022},
}

@article{meyers2009use,
	title = {How to use (five) curriculum design principles to align authentic learning environments, assessment, students’ approaches to thinking and learning outcomes},
	author = {Meyers, Noel M and Nulty, Duncan D},
	journal = {Assessment \& Evaluation in Higher Education},
	number = {5},
	pages = {565--577},
	publisher = {Taylor \& Francis},
	volume = {34},
	year = {2009},
}

@article{musa2012project,
	title = {Project-based learning (PjBL): Inculcating soft skills in 21st century workplace},
	author = {Musa, Faridah and Mufti, Norlaila and Latiff, Rozmel Abdul and Amin, Maryam Mohamed},
	journal = {Procedia-Social and Behavioral Sciences},
	pages = {565--573},
	publisher = {Elsevier},
	volume = {59},
	year = {2012},
}

@article{arafeh2016curriculum,
	title = {Curriculum mapping in higher education: a case study and proposed content scope and sequence mapping tool},
	author = {Arafeh, Sousan},
	journal = {Journal of Further and Higher Education},
	number = {5},
	pages = {585--611},
	publisher = {Taylor \& Francis},
	volume = {40},
	year = {2016},
}

@article{tian2024enhancing,
	title = {Enhancing instructional quality: Leveraging computer-assisted textual analysis to generate in-depth insights from educational artifacts},
	author = {Tian, Zewei and Sun, Min and Liu, Alex and Sarkar, Shawon and Liu, Jing},
	journal = {arXiv preprint arXiv:2403.03920},
	year = {2024},
}

@inproceedings{durant2015ce2016,
  title={CE2016: Updated computer engineering curriculum guidelines},
  author={Durant, Eric and Impagliazzo, John and Conry, Susan and Reese, Robert and Lam, Herman and Nelson, Victor and Hughes, Joseph and Liu, Weidong and Lu, Junlin and McGettrick, Andrew},
  booktitle={2015 IEEE Frontiers in Education Conference (FIE)},
  pages={1--2},
  year={2015},
  doi = {10.1109/FIE.2015.7344157},
  organization={IEEE}
}

\newpage
\appendix
\section*{Appendix}

\section{Details of competency frameworks}

\begin{table}[H]
    \centering
    \caption{Summary of competency items}\label{tab:Summary of competency items} 
    \begin{tabular}{p{4cm} p{11cm}} 
    \toprule
          Competency frameworks& Competencies\\ 
        \midrule
         O*NET(Workforce Competencies)& Complex problem solving; Critical thinking; Deductive reasoning; Judgment and decision making; Inductive reasoning; Category flexibility; Perceptual speed; Information ordering; Cooperation; Social Interaction; Concerns for others; Leadership; Persistence; Achievement/effort; Initiative; Originality; Innovation; Oral expression; Oral comprehension; Written expression; Written comprehension
         \\ 
               EU(Key Competences)& Literacy competence; Multilingual competence; Mathematical competence and competence in science, technology and engineering; Digital competence; Personal, social and learning to learn competence; Citizenship competence; Entrepreneurship competence; Cultural awareness and expression competence
               \\
            
                    ESDC (Success Model)& Adaptability; Collaboration; Communication; Creativity and Innovation; Digital; Numeracy; Problem Solving; Reading; Writing\\
\bottomrule
    \end{tabular}
\end{table}

\section{Human annotation} \label{Human annotation process}

\subsection{Score distribution in human annotation}
\begin{table}[H]
\scriptsize
\centering
\caption{Score distribution across competency frameworks}
\resizebox{\textwidth}{!}{%
\begin{tabularx}{\textwidth}{l c *{5}{>{\centering\arraybackslash}p{1cm} >{\centering\arraybackslash}p{1cm}}}
\toprule
\makecell{Competency\\frameworks} & Score &
\multicolumn{2}{c}{\makecell{Learning\\Objective}} &
\multicolumn{2}{c}{\makecell{Concise\\course\\description}} &
\multicolumn{2}{c}{\makecell{Detailed\\course\\description}} &
\multicolumn{2}{c}{\makecell{Learning\\activity\\content}} &
\multicolumn{2}{c}{\makecell{Instructional\\schedule}} \\
\cmidrule(lr){3-12}
& & N & \% & N & \% & N & \% & N & \% & N & \% \\
\midrule
EU & 0 & 106 & 30.8 & 157 & 40.9 & 142 & 38.6 & 113 & 35.3 & 100 & 31.2 \\
   & 1 & 57 & 16.6 & 42 & 10.9 & 50 & 13.6 & 55 & 17.2 & 45 & 14.1 \\
   & 2 & 57 & 16.6 & 54 & 14.1 & 50 & 13.6 & 55 & 17.2 & 30 & 9.4 \\
   & 3 & 62 & 18.0 & 47 & 12.2 & 48 & 13.0 & 78 & 24.4 & 60 & 18.8 \\
   & NA & 62 & 18.0 & 84 & 21.9 & 78 & 21.2 & 19 & 5.9 & 85 & 26.6 \\
\midrule
O*NET & 0 & 50 & 5.5 & 126 & 12.5 & 128 & 13.3 & 21 & 2.5 & 182 & 21.7 \\
      & 1 & 201 & 22.3 & 275 & 27.3 & 219 & 22.7 & 213 & 25.4 & 111 & 13.2 \\
      & 2 & 373 & 41.3 & 286 & 28.4 & 320 & 33.1 & 319 & 38.0 & 164 & 19.5 \\
      & 3 & 92 & 10.2 & 53 & 5.3 & 67 & 6.9 & 78 & 9.3 & 5 & 0.6 \\
      & NA & 187 & 20.7 & 268 & 26.6 & 232 & 24.0 & 209 & 24.9 & 378 & 45.0 \\
\midrule
ESDC & 0 & 33 & 9.6 & 47 & 12.2 & 54 & 14.7 & 55 & 17.2 & 86 & 26.9 \\
     & 1 & 99 & 28.8 & 132 & 34.4 & 77 & 20.9 & 85 & 26.6 & 54 & 16.9 \\
     & 2 & 121 & 35.2 & 110 & 28.6 & 127 & 34.5 & 132 & 41.2 & 64 & 20.0 \\
     & 3 & 50 & 14.5 & 44 & 11.5 & 34 & 9.2 & 18 & 5.6 & 2 & 0.6 \\
     & NA & 41 & 11.9 & 51 & 13.3 & 76 & 20.7 & 30 & 9.4 & 114 & 35.6 \\
\bottomrule
\end{tabularx}%
}
\label{annotation_distribution}
\end{table}

\subsection{human annotation notes}
\begin{table}[H]
\small
    \centering
    \caption{O*NET(Workforce Competencies)}\label{O*NET(Workforce Competencies)} 
    \begin{tabular}{p{4cm} p{12cm} } 
    \toprule
          Competency & Definition\\ 
        \midrule
         Complex problem solving & Identifying complex problems and reviewing related information to develop and evaluate options and implement solutions.\\ 
               Critical thinking& Using logic and reasoning to identify the strengths and weaknesses of alternative solutions, conclusions, or approaches to problems.\\
              Deductive reasoning& Applying general rules to specific situations to produce logical answers, ensuring that the reasoning process is transparent and justifiable. \\
              Judgment and decision making& Evaluating the relative costs and benefits of various actions or solutions, and choosing the option that is most appropriate for the context.\\
              Inductive reasoning& Integrating multiple pieces of information to formulate general rules or conclusions, including identifying patterns or relationships among seemingly unrelated data.\\
              Category flexibility& Generating or applying different sets of rules to combine or group things in novel ways, and flexibly adapting categorization criteria to fit new situations or information.\\
              Perceptual speed& Quickly and accurately comparing similarities and differences among sets of letters, numbers, objects, pictures, or patterns, and efficiently identifying discrepancies or matches.\\
              Information ordering& Arranging actions, information, or objects in a specific sequence according to established rules, instructions, or logical patterns, ensuring that each step follows a coherent and justifiable order.\\
              Cooperation& Demonstrating a pleasant and cooperative attitude with others in the work environment, actively contributing to group efforts, and supporting colleagues to achieve common goals.\\
              Social interaction / Social orientation& Preferring to work collaboratively with others rather than independently, establishing and maintaining effective interpersonal relationships, and engaging in constructive social exchanges within a group or team.\\
              Concern for others& Showing sensitivity to the needs and feelings of others, offering understanding and practical help, and responding appropriately to others’ emotional or situational challenges.\\
              Leadership& Willingly taking charge of situations or groups, offering direction and constructive opinions, motivating others, and assuming responsibility for group outcomes or decisions.\\
              Persistence& Sustaining effort and motivation in the face of obstacles, setbacks, or difficulties, continuing to pursue goals despite challenges or repeated failures.\\
              Achievement/effort& Setting personally challenging achievement goals, maintaining high standards for one’s own performance, and exerting consistent effort to master difficult tasks.\\
              Initiative& Proactively taking on new responsibilities and challenges, identifying and acting on opportunities without needing external prompting or supervision.\\
              Originality& Developing unusual or clever ideas for a given topic, problem, or situation, and producing creative solutions or approaches that differ from conventional practices.\\
              Innovation& Formulating and implementing novel ideas, processes, or products that improve outcomes or create added value, and actively seeking opportunities for creative change.\\
              Oral expression& Communicating information and ideas clearly and effectively in spoken language, ensuring that listeners can readily understand the intended meaning.\\
              Oral comprehension& Listening to and accurately understanding information, instructions, or ideas presented through spoken words and sentences in various contexts.\\
              Written expression& Conveying information and ideas effectively in writing, producing clear and understandable documents, messages, or reports tailored to the audience.\\
              Written comprehension& Reading and comprehending information and ideas presented in written form, accurately interpreting meaning, context, and relevant details.\\
\bottomrule
    \end{tabular}
\end{table}

\begin{table}[H]
    \centering
    \caption{EU(Key Competences)}\label{EU(Key competences)} 
    \begin{tabular}{p{4cm} p{12cm} } 
    \toprule
          Competency & Definition\\ 
        \midrule
         Literacy competence & The ability to identify, understand, express, create, and interpret concepts, feelings, facts, and opinions in both oral and written forms, using visual, audio, and digital materials across disciplines and contexts.\\ 
               Multilingual competence& The ability to use different languages appropriately and effectively for communication.\\
              Mathematical competence and competence in science, technology, and engineering& The ability to develop and apply mathematical thinking and understanding to solve a range of problems in everyday situations. Competence in science, technology, and engineering involves the application of knowledge and methodology to explain the natural world and to use that knowledge to identify questions and draw evidence-based conclusions. \\
              Digital competence& The confident, critical, and responsible use of digital technologies for learning, at work, and for participation in society.\\
              Personal, social, and learning to learn competence& The ability to reflect upon oneself, manage time and information effectively, work with others in a constructive way, remain resilient, and manage one’s own learning and career.\\
              Citizenship competence& The ability to act as responsible citizens and to fully participate in civic and social life, based on understanding of social, economic, legal, and political concepts and structures, as well as global developments and sustainability.\\
              Entrepreneurship competence& The capacity to act upon opportunities and ideas, and to transform them into values for others.\\
               Cultural awareness and expression competence& The understanding of and respect for how ideas and meaning are creatively expressed and communicated in different cultures and through a range of arts and other cultural forms.\\
\bottomrule
    \end{tabular}
\end{table}

\begin{table}[H]
\small
    \centering
    \caption{ESDC (Success Model)}\label{ESDC (Success model)} 
    \begin{tabular}{p{4cm} p{12cm} } 
    \toprule
          Competency & Definition\\ 
        \midrule
         Adaptability & Adjusting behavior or strategies in response to change or unexpected challenges, while maintaining focus and persistence.\\ 
               Collaboration& Working constructively with others to achieve shared goals, including communicating openly and sharing responsibilities.\\
              Communication& Expressing and understanding information clearly through various modes, and adapting messages appropriately for the context. \\
              Communication& Expressing and understanding information clearly through various modes, and adapting messages appropriately for the context.\\
              Creativity and Innovation& Generating and applying new ideas or approaches to improve processes, solve problems, or create value.\\
              Digital& Effectively using digital tools and technologies to find, manage, and communicate information.\\
              Numeracy& Interpreting and applying quantitative information and calculations to solve practical problems.\\
              Problem Solving& Identifying issues, analyzing relevant information, and implementing effective solutions.\\
              Reading& Understanding, interpreting, and using written information to accomplish tasks.\\
              Writing& Organizing and conveying information clearly and effectively in written form.\\
              
\bottomrule
    \end{tabular}
\end{table}

\section{Comprehensive evaluation results}
\subsection{RQ2}
\begin{table}[H]
\small
    \centering
    \caption{Zero-shot performance across different levels of classification granularity. }\label{tab:rq2 All performance metrics} 
    \begin{tabular}{llllllll} 
    \toprule
          $N_{class}$& Model& Kappa& ICC& Accuracy& Precision&Recall& F1\\ 
        \midrule
         5& Llama3-8B& 0.248& 0.396& 0.299& 0.312&0.274& 0.365\\
              5& Llama3-70B& 0.304& 0.474& 0.35& 0.338&0.309& 0.383\\
 5& GPT 3.5-turbo& 0.282& 0.429& 0.345& 0.294& 0.272&0.39\\
              5& GPT4o& 0.301& 0.456& 0.386& 0.362&0.31& 0.416\\
     \midrule
              4& Llama3-8B& 0.204& 0.343& 0.44& 0.406&0.34& 0.48\\
              4& Llama3-70B& 0.246& 0.386& 0.498& 0.442&0.376& 0.52\\
 4& GPT 3.5-turbo& 0.244& 0.369& 0.472& 0.401& 0.345&0.49\\
              4& GPT4o& 0.221& 0.333& 0.559& 0.491&0.366& 0.575\\
  \midrule
 3& Llama3-8B& 0.204& 0.343& 0.44& 0.406& 0.34&0.48\\
 3& Llama3-70B& 0.258& 0.413& 0.549& 0.505& 0.476&0.528\\
 3& GPT 3.5-turbo& 0.25& 0.396& 0.529& 0.474& 0.447&0.508\\
 3& GPT4o& 0.224& 0.351& 0.605& 0.55& 0.467&0.588\\
   \midrule
 2& Llama3-8B& 0.213& 0.319& 0.715& 0.682& 0.618&0.679\\
 2& Llama3-70B& 0.178& 0.264& 0.724& 0.701& 0.601&0.721\\
 2& GPT 3.5-turbo& 0.192& 0.267& 0.724& 0.661& 0.602&0.704\\
 2& GPT4o& 0.155& 0.231& 0.729& 0.715& 0.593&0.742\\
 \bottomrule
    \end{tabular}
\end{table}

\subsection{RQ3}
\begin{table}[H]
\small
    \centering
    \caption{Zero-Shot with Competency Definition (DEF) performance across different levels of classification granularity. }\label{tab:rq2 All performance metrics} 
    \begin{tabular}{llllllll} 
    \toprule
          $N_{class}$& Model& Kappa& ICC& Accuracy& Precision&Recall& F1\\ 
        \midrule
         5& Llama3-8B& 0.239& 0.395& 0.305& 0.311&0.269& 0.283\\
              5& Llama3-70B& 0.345& 0.521& 0.337& 0.316&0.286& 0.311\\
 5& GPT 3.5-turbo& 0.254& 0.393& 0.337& 0.29& 0.275&0.285\\
              5& GPT4o& 0.338& 0.515& 0.363& 0.331&0.282& 0.344\\
     \midrule
              4& Llama3-8B& 0.239& 0.395& 0.305& 0.311&0.269& 0.283\\
              4& Llama3-70B& 0.345& 0.521& 0.337& 0.316&0.286& 0.311\\
 4& GPT 3.5-turbo& 0.254& 0.393& 0.337& 0.29& 0.275&0.285\\
              4& GPT4o& 0.338& 0.515& 0.363& 0.331&0.282& 0.344\\
  \midrule
 3& Llama3-8B& 0.189& 0.327& 0.455& 0.389& 0.322&0.389\\
 3& Llama3-70B& 0.298& 0.474& 0.533& 0.487& 0.467&0.445\\
 3& GPT 3.5-turbo& 0.212& 0.333& 0.512& 0.436& 0.425&0.417\\
 3& GPT4o& 0.248& 0.406& 0.571& 0.495& 0.444&0.49\\
   \midrule
 2& Llama3-8B& 0.206& 0.304& 0.7& 0.637& 0.597&0.597\\
 2& Llama3-70B& 0.2& 0.291& 0.701& 0.666& 0.596&0.652\\
 2& GPT 3.5-turbo& 0.129& 0.171& 0.671& 0.578& 0.558&0.577\\
 2& GPT4o& 0.147& 0.227& 0.69& 0.682& 0.579&0.687\\
 \bottomrule
    \end{tabular}
\end{table}

\begin{table}[H]
\small
    \centering
    \caption{Curricular CoT (CQA) performance across different levels of classification granularity. }\label{tab:rq2 All performance metrics} 
    \begin{tabular}{llllllll} 
    \toprule
          $N_{class}$& Model& Kappa& ICC& Accuracy& Precision&Recall& F1\\ 
        \midrule
         5& Llama3-8B& 0.304& 0.473& 0.323& 0.303&0.286& 0.299\\
              5& Llama3-70B& 0.337& 0.519& 0.338& 0.338&0.292& 0.319\\
 5& GPT 3.5-turbo& 0.304& 0.481& 0.322& 0.291& 0.279&0.281\\
              5& GPT4o& 0.35& 0.505& 0.378& 0.375&0.292& 0.351\\
     \midrule
              4& Llama3-8B& 0.252& 0.416& 0.452& 0.385&0.344& 0.4\\
              4& Llama3-70B& 0.279& 0.444& 0.484& 0.433&0.357& 0.444\\
 4& GPT 3.5-turbo& 0.261& 0.415& 0.444& 0.367& 0.34&0.366\\
              4& GPT4o& 0.264& 0.403& 0.536& 0.48&0.348& 0.465\\
  \midrule
 3& Llama3-8B& 0.252& 0.416& 0.452& 0.385& 0.344&0.4\\
 3& Llama3-70B& 0.299& 0.477& 0.551& 0.505& 0.48&0.46\\
 3& GPT 3.5-turbo& 0.277& 0.45& 0.535& 0.491& 0.448&0.43\\
 3& GPT4o& 0.273& 0.423& 0.599& 0.546& 0.464&0.503\\
   \midrule
 2& Llama3-8B& 0.165& 0.259& 0.704& 0.673& 0.585&0.703\\
 2& Llama3-70B& 0.183& 0.281& 0.701& 0.655& 0.589&0.645\\
 2& GPT 3.5-turbo& 0.209& 0.308& 0.712& 0.634& 0.595&0.605\\
 2& GPT4o& 0.251& 0.372& 0.71& 0.663& 0.623&0.637\\
 \bottomrule
    \end{tabular}
\end{table}

\begin{table}[H]
\small
    \centering
    \caption{Curricular CoT (CQ) performance across different levels of classification granularity. }\label{tab:rq2 All performance metrics} 
    \begin{tabular}{llllllll} 
    \toprule
          $N_{class}$& Model& Kappa& ICC& Accuracy& Precision&Recall& F1\\ 
        \midrule
         5& Llama3-8B& 0.279& 0.444& 0.316& 0.29&0.278& 0.293\\
              5& Llama3-70B& 0.332& 0.516& 0.334& 0.342&0.289& 0.31\\
 5& GPT 3.5-turbo& 0.288& 0.445& 0.321& 0.273& 0.274&0.293\\
              5& GPT4o& 0.35& 0.519& 0.37& 0.35&0.287& 0.363\\
     \midrule
              4& Llama3-8B& 0.232& 0.389& 0.448& 0.37&0.339& 0.385\\
              4& Llama3-70B& 0.273& 0.432& 0.479& 0.435&0.352& 0.429\\
 4& GPT 3.5-turbo& 0.254& 0.399& 0.438& 0.33& 0.335&0.377\\
              4& GPT4o& 0.252& 0.393& 0.531& 0.449&0.338& 0.487\\
  \midrule
 3& Llama3-8B& 0.232& 0.389& 0.448& 0.37& 0.339&0.385\\
 3& Llama3-70B& 0.285& 0.463& 0.541& 0.498& 0.468&0.455\\
 3& GPT 3.5-turbo& 0.265& 0.435& 0.517& 0.458& 0.439&0.422\\
 3& GPT4o& 0.258& 0.412& 0.584& 0.52& 0.454&0.502\\
   \midrule
 2& Llama3-8B& 0.262& 0.388& 0.717& 0.657& 0.626&0.633\\
 2& Llama3-70B& 0.19& 0.291& 0.705& 0.672& 0.593&0.647\\
 2& GPT 3.5-turbo& 0.197& 0.291& 0.706& 0.635& 0.59&0.58\\
 2& GPT4o& 0.165& 0.256& 0.704& 0.688& 0.586&0.695\\
 \bottomrule
    \end{tabular}
\end{table}

\begin{table}[H]
\small
    \centering
    \caption{Curricular CoT (QA) performance across different levels of classification granularity. }\label{tab:rq2 All performance metrics} 
    \begin{tabular}{llllllll} 
    \toprule
          $N_{class}$& Model& Kappa& ICC& Accuracy& Precision&Recall& F1\\ 
        \midrule
         5& Llama3-8B& 0.277& 0.43& 0.307& 0.298&0.273& 0.288\\
              5& Llama3-70B& 0.318& 0.483& 0.341& 0.322&0.291& 0.324\\
 5& GPT 3.5-turbo& 0.291& 0.454& 0.328& 0.261& 0.269&0.3\\
              5& GPT4o& 0.334& 0.491& 0.375& 0.364&0.293& 0.349\\
     \midrule
              4& Llama3-8B& 0.231& 0.381& 0.437& 0.373&0.337& 0.396\\
              4& Llama3-70B& 0.272& 0.419& 0.487& 0.424&0.357& 0.459\\
 4& GPT 3.5-turbo& 0.245& 0.387& 0.446& 0.323& 0.325&0.393\\
              4& GPT4o& 0.256& 0.394& 0.535& 0.461&0.347& 0.468\\
  \midrule
 3& Llama3-8B& 0.231& 0.381& 0.437& 0.373& 0.337&0.396\\
 3& Llama3-70B& 0.284& 0.445& 0.55& 0.501& 0.476&0.48\\
 3& GPT 3.5-turbo& 0.253& 0.416& 0.525& 0.445& 0.426&0.421\\
 3& GPT4o& 0.267& 0.42& 0.594& 0.528& 0.463&0.497\\
   \midrule
 2& Llama3-8B& 0.287& 0.435& 0.706& 0.673& 0.645&0.634\\
 2& Llama3-70B& 0.197& 0.305& 0.705& 0.663& 0.595&0.657\\
 2& GPT 3.5-turbo& 0.188& 0.278& 0.707& 0.625& 0.583&0.596\\
 2& GPT4o& 0.184& 0.299& 0.705& 0.661& 0.595&0.643\\
 \bottomrule
    \end{tabular}
\end{table}

\begin{table}[H]
\small
    \centering
    \caption{Curricular CoT (A) performance across different levels of classification granularity. }\label{tab:rq2 All performance metrics} 
    \begin{tabular}{llllllll} 
    \toprule
          $N_{class}$& Model& Kappa& ICC& Accuracy& Precision&Recall& F1\\ 
        \midrule
         5& Llama3-8B& 0.261& 0.42& 0.304& 0.308&0.271& 0.289\\
              5& Llama3-70B& 0.319& 0.481& 0.347& 0.324&0.292& 0.327\\
 5& GPT 3.5-turbo& 0.29& 0.462& 0.332& 0.295& 0.268&0.278\\
              5& GPT4o& 0.347& 0.503& 0.381& 0.361&0.299& 0.347\\
     \midrule
              4& Llama3-8B& 0.218& 0.368& 0.424& 0.384&0.327& 0.392\\
              4& Llama3-70B& 0.269& 0.413& 0.493& 0.414&0.352& 0.454\\
 4& GPT 3.5-turbo& 0.252& 0.406& 0.454& 0.374& 0.323&0.356\\
              4& GPT4o& 0.268& 0.407& 0.538& 0.454&0.347& 0.451\\
  \midrule
 3& Llama3-8B& 0.218& 0.368& 0.424& 0.384& 0.327&0.392\\
 3& Llama3-70B& 0.281& 0.439& 0.557& 0.501& 0.469&0.463\\
 3& GPT 3.5-turbo& 0.265& 0.438& 0.533& 0.468& 0.436&0.424\\
 3& GPT4o& 0.279& 0.431& 0.599& 0.528& 0.465&0.513\\
   \midrule
 2& Llama3-8B& 0.263& 0.405& 0.696& 0.668& 0.631&0.64\\
 2& Llama3-70B& 0.204& 0.317& 0.707& 0.655& 0.597&0.641\\
 2& GPT 3.5-turbo& 0.213& 0.336& 0.712& 0.662& 0.596&0.591\\
 2& GPT4o& 0.192& 0.303& 0.71& 0.67& 0.599&0.663\\
 \bottomrule
    \end{tabular}
\end{table}

\end{document}